\documentclass[aps,pra,reprint,showpacs,floatfix,superscriptaddress]{revtex4-1}

\usepackage{amssymb,amsmath,amstext}                %%   American Physical Society math etc extensions
\usepackage{graphicx}                                               %%   Include figure files                                            
\usepackage{epstopdf}                                               %%   help with eps -> pdf 
\usepackage{color}                                                     %%   change color
\usepackage{bm}                                                        %%   bold math                                    
\usepackage{appendix}                                              %%   formating appendicies
\usepackage[utf8]{inputenc}
\usepackage{bbold}
\usepackage{bbm}
\usepackage{latexsym}
\usepackage[T1]{fontenc}
\usepackage{xcolor}
\usepackage{braket}
\definecolor{lblue} {RGB}{51,71,158}
\usepackage[colorlinks=true,citecolor=blue,linkcolor=blue,urlcolor=lblue]{hyperref}

\usepackage[normalem]{ulem}

\definecolor{darkgreen}{rgb}{0.13, 0.55, 0.13}

\def\beq{\begin{equation}}
\def\eeq{\end{equation}}

\begin{document}

\title{Bond order via cavity-mediated interactions}
\author{Titas Chanda}
\affiliation{The Abdus Salam International Centre for Theoretical Physics (ICTP), Strada Costiera 11, 34151 Trieste, Italy}
\affiliation{Institute of Theoretical Physics, Jagiellonian University in Krakow, \L{}ojasiewicza 11,
	30-348 Krak\'ow, Poland}
\author{Rebecca Kraus}
\affiliation{Theoretical Physics, Department of Physics, Saarland University, 66123 Saarbr\"ucken, Germany}
\author{Jakub Zakrzewski}
\affiliation{Institute of Theoretical Physics, Jagiellonian University in Krakow, \L{}ojasiewicza 11,
	30-348 Krak\'ow, Poland}
\affiliation{Mark Kac Complex Systems Research Center, Jagiellonian University in Krak\'ow, \L{}ojasiewicza 11, 30-348 Krak\'ow, Poland}
\author{Giovanna Morigi}
\affiliation{Theoretical Physics, Department of Physics, Saarland University, 66123 Saarbr\"ucken, Germany}

\begin{abstract}
We numerically study the phase diagram of bosons tightly trapped in the lowest band of an optical lattice and dispersively coupled to a single-mode cavity field. The dynamics is encompassed by an extended Bose-Hubbard model. Here, the cavity-mediated interactions are described by a two-body potential term with a global range and by a correlated tunneling term where the hopping amplitude depends on a global observable. We determine the ground state properties in one dimension by means of the density matrix renormalization group algorithm, focusing  on the effects due to the correlated tunneling. The latter is responsible for the onset of bond orders, manifesting {in} one insulating and two gapless bond ordered phases. 
We discuss the resulting phases for different geometries that correspond to different relative strengths of the correlated tunneling coefficient. 
We finally analyze the scaling of the entanglement entropy in the gapless bond ordered phases that appear entirely due to global interactions and determine the corresponding central charges.
\end{abstract}
\date{\today} % TODO: update
\maketitle

\section{Introduction}

Ultracold atomic gases in optical lattices realize the strongly-correlated dynamics of the Hubbard model with tunable interactions \cite{Fisher1989,Jaksch98,Greiner2002,Bloch08,Lewenstein12}.
The dispersive coupling with a high-finesse resonator, additionally, allows one to design interactions whose range can be tailored and {whose} strength can be tuned \cite{Gopalakrishnan11,Schleier-Smith21}. One prominent example is the all-to-all interaction in a quantum gas of bosons {realized by coupling {an electric dipole} transition with a single-mode resonator} \cite{Landig16}. {Here, t}he appearance of phases with  density modulations was observed  by tuning the {effective strength of the coupling with the cavity}. These patterns support coherent scattering into the cavity mode \cite{Landig16} and can be either superfluid or incompressible. The experimentally measured phases are captured by the ground state of an extended Bose-Hubbard model with global interactions, as shown in \cite{Dogra16, Niederle16, Flottat17, Himbert19}. {This model describes} the effect of the cavity{-mediated}  potential by means of an interaction term between pair of sites that depends on the {onsite} density and has a global range. 

{\it Ab-initio} derivations of the Bose-Hubbard model show, however, that two-body interactions give rise to {further} terms {describing} correlated tunneling \cite{Fernandez10,Maik2013,Dutta15,Elliott16,Caballero16,Chanda21}. These contributions can be important in determining the phase of superconductors \cite{Strack93,Hirsch94,Amadon96}, frustrated quantum magnets \cite{Schmidt08,Schmidt06}, and dipolar gases \cite{Sowinski12,Maik2013,Dutta15,Biedron18,Kraus20}. They can interfere with single-particle hopping \cite{Kraus20,Suthar20,Chanda21} and, in the cavity {quantum electrodynamics} (QED) setup, give rise to an effective periodic modulation of the bonds. {Indeed, a}t half filling this interference is at the basis of the emergence of self-organized topological insulating phases \cite{Chanda21}.

In this work, we characterize the quantum ground state of the extended Bose-Hubbard model of cavity QED taking into account both the cavity-induced density-density and correlated hopping terms for different lattice geometries. We consider a one-dimensional lattice and determine the quantum phases for  half and  unit fillings using the density matrix renormalization group (DMRG)  algorithm \cite{White92,White93,Schollwoeck11, Orus14} extended to the case of global interactions \cite{Chanda20,Chanda21}. We show that correlated tunneling gives rise to a bond order that can be either supersolid, superfluid, or insulating. We analyze, in particular, the scaling of the entanglement entropy at the bond superfluid and bond supersolid phases. Our analysis complements and extends the study of  \cite{Chanda21} by investigating the phase diagram for generic geometries. These geometries were also considered in \cite{Caballero16} where the ground state of atoms in small chains using exact diagonalization was {discussed}.

The paper is organized as follows. In Sec.~\ref{Sec:2} we review the extended Bose-Hubbard model of cavity QED and discuss the dependence of its coefficients on the cavity system parameters. In Sec.~\ref{Sec:4}, we present the ground state phase diagrams calculated by means of the DMRG approach and analyze the nature of the phase transitions. We {then} perform the scaling analysis of the entanglement entropy in the gapless bond ordered phases that are stabilized entirely by the cavity-mediated global terms.
The conclusions are drawn in Sec.~\ref{Sec:5}. The Appendix provides details on the determination of the coefficients of the extended Bose-Hubbard model.

\section{Extended Bose-Hubbard model with cavity-mediated interactions}
\label{Sec:2}

The system we consider consists of $N$ bosons tightly bound in the lowest band of an one dimensional optical lattice with $L$ sites. Let $\hat a_i$ and $\hat a_i^\dagger$ denote the bosonic operators destroying and creating, respectively, a boson at site $i=1,\ldots,L$ with $\left[\hat{a}_{i}, \hat{a}^\dagger_{j}\right]=\delta_{i,j}$, and $\hat n_i=\hat a_i^\dagger\hat a_i$ being the corresponding particle number operator. The Hamiltonian $\hat{H}_{\text{EBH}}$ determining their dynamics is the one of the extended Bose-Hubbard model \cite{Habibian13,Caballero16}:
\begin{equation}
\label{Eq:BH}
\hat{H}_{\text{EBH}}=\hat{H}_{\text{BH}}+\hat{H}_{\text{BH}}^{C}\,,
\end{equation}
which is the sum of the standard Bose-Hubbard Hamiltonian $\hat{H}_{\text{BH}}$,
\begin{align}
\label{Eq:BH:0}
	\hat{H}_{\text{BH}}  = -t \sum_j \left( \hat{a}^{\dagger}_j \hat{a}_{j+1} + \text{H.c.}\right) + \frac{U}{2} \sum_j \hat{n}_j (\hat{n}_j -1)\,,
	\end{align}
and of the terms containing the cavity-mediated interactions \cite{Fernandez10,Caballero16}:
\begin{align}
\hat{H}_{\text{BH}}^{C}\,=\frac{U_1}{L}\left(z^2\hat{D}^2 +  zy\left(\hat{D}\hat{B}+\hat{B}\hat{D} \right) +y^2\hat{B}^2 \right)\,.  \label{LRBH_alphaone}
\end{align}
The details of the derivation of Hamiltonian \eqref{LRBH_alphaone} follow from \cite{Habibian13,Habibian13b,Niederle16} and are reported in  \cite{Chanda21} (see also the Appendix \ref{app:BH}). The coefficients in Eq. \eqref{Eq:BH:0} are positive and are the nearest-neighbor hopping rate $t$ and the strength of the onsite repulsion $U$. The coefficient $U_1$ scaling the cavity term in Eq. \eqref{LRBH_alphaone} can be either positive or negative, the factor $1/L$  warrants extensivity \cite{Fernandez10,Habibian13}.  
Operators $\hat{D}$ and $\hat{B}$ depend on the onsite densities and hoppings, respectively \cite{Caballero16,Chanda21}:
	\begin{align}
		&\hat{D}=\sum_{j}(-1)^{j}\hat{n}_{j}   \label{Dfield}\,,\\
		&\hat{B}=\sum_i(-1)^{i}\left(\hat{a}^\dagger_{i+1}\hat{a}_{i}+{\rm H.c.}\right) \,,
		\label{cavity_field}
	\end{align}  
where the staggered sum emerges when the {cavity wavelength is twice the lattice periodicity} \cite{Habibian13,Niederle16}. The coefficients $y$ and $z$ are dimensionless parameters whose strength depends on the setup's geometry and are discussed in the following. 

We note that the term $\hat D^2$ is a global density-density interaction, that promotes the onset of a population imbalance between even sites (with $j=2n$) and odd sites (with $j=2n+1$)  \cite{Landig16,Dogra16,Niederle16}. The two other terms,
$\hat{D}\hat{B}+\hat{B}\hat{D}$ and $\hat B^2$, describe correlated tunneling processes induced by the cavity field. %Terms describing pair tunnelling have not been included in  \eqref{LRBH_alphaone}  because they are typically much smaller in amplitude. 

\subsection{Bose-Hubbard coefficients}

The coefficients in Eq.\ \eqref{Eq:BH} are numerically calculated from the overlap integrals using the Wannier functions of the lowest band of the static optical lattice. In our calculations they are varied taking into account that in the experiment the control parameters are the depth of the optical lattice, the $s$-wave scattering length, and the cavity interaction amplitude and {its} sign. When we sweep across the phase diagram, we keep fixed the lattice depth at the value $V_0=4E_R$, where $E_R=\hbar^2k^2/(2m)$ is the recoil energy for atoms of mass $m$ and $k$ the lattice and cavity wave number. Therefore, in our {numerical} calculations  the tunneling coefficient is kept constant. The ratio $t/U$ is varied by tuning $U$ via the $s$-wave scattering length. The sign of the detuning between cavity and driving laser determines the sign of the coefficient $U_1$. Moreover, the detuning and the strength of the cavity field determine the magnitude of $|U_1|$\cite{Baumann10,Fernandez10,Habibian13,Habibian13b}. Thus, in our calculations the ratio $U_1/U$ and $t/U$ are varied by simultaneously changing $U_1$ and $U$.  

The coefficients $y$ and $z$ in Eq.\ \eqref{LRBH_alphaone} are overlap integrals between the Wannier functions $w_j(x)$ and the cavity mode function $\cos(kx+\phi)$ (see Appendix \ref{app:BH}):
	\begin{align}
		z= \int_0^{aL} dx & \ w_j(x)^2\cos(k x+\phi) \nonumber \\ y= \int_0^{aL} dx & \ w_j (x)w_{j+1}(x)\cos( k x+\phi)  \ ,  \label{overlap}
	\end{align}
where $j$ denotes the lattice site about which the Wannier function is centered, $a$ is the lattice periodicity, $a=\pi/k$, and $\phi$ is the phase shift between the lattice and the cavity standing wave. The phase shift $\phi=0$  and $\phi=\pi/2$ correspond to trapping the atoms at the antinodes and at the nodes, respectively, of the cavity standing wave. This is realized by either tuning the laser on the blue or on the red side of the cavity resonance. In this paper we will also consider the case $\phi=\pi/4$, for which both $y$ and $z$ are different from zero. We remark that $y\le 0$ for the parameter regimes we inspect.  

\begin{figure}
		\includegraphics[width=0.8\linewidth]{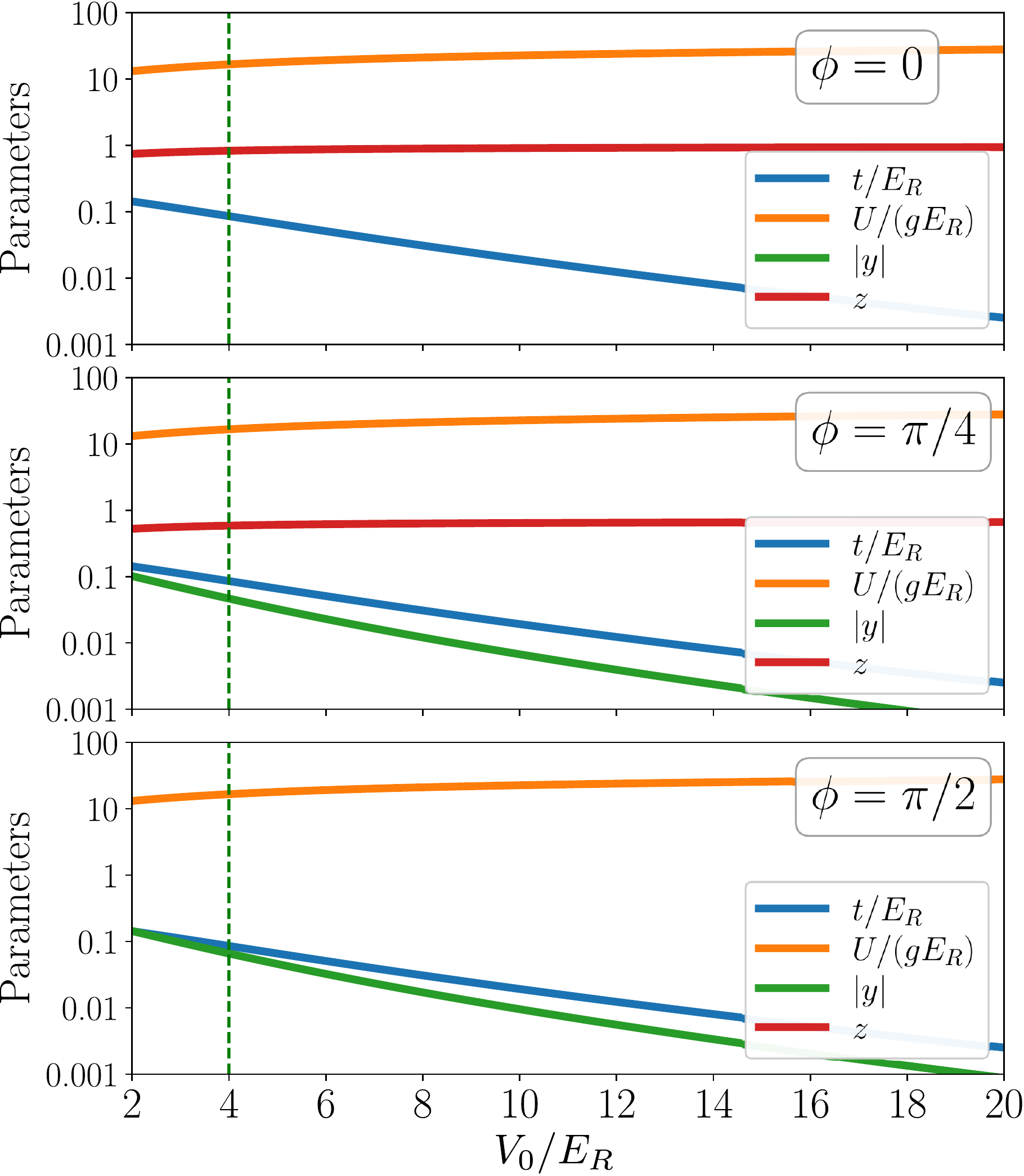}
		\caption{(Color online.) The coefficients $z$, $|y|$, $t/E_R$ and $U/(gE_R)$ as a function of the lattice depth $V_0$, in units of the recoil energy $E_R$, for $\phi=0$ (upper panel), $\phi=\pi/4$ (middle panel), and $\phi=\pi/2$ (lower panel). The vertical dashed line corresponds to the lattice depth of $V_0= 4E_R$. The coefficient $g$ scaling the onsite repulsion coefficient contains the physical variables including the scattering length, therefore $U/(gE_R)$ is the overlap integral solely depending on the Wannier functions, which in turn are determined by $V_0$.
		}
		\label{Fig1}
	\end{figure}

Figure~\ref{Fig1} displays the coefficients $z$ and $|y|$ as a function of $V_0$ for the three different phase shifts $\phi=0,\pi/4,\pi/2$ considered in this paper. We also display the tunneling rate $t$ and the on-site coefficient $U$ for comparison, keeping in mind that these quantities are independent of $\phi$. The upper panel shows the parameters for $\phi=0$, where the $y$ coefficient vanishes within machine precision. For $\phi= \pi/4$ both $y$ and $z$ are finite:  the $z$ coefficient is almost independent of the lattice depth $V_0$, while $|y|$ decreases monotonically with $V_0$. For $\phi= \pi/2$ (see lower panel) the $z$ coefficient is zero within machine precision. The vertical dashed line in Fig.~\ref{Fig1} indicates the value of $V_0$ considered in this work. For $\phi=0$ the value of the overlap integrals are given by $z=0.8279$ and $y=0$, thus we neglect the correlated tunneling. For $\phi= \pi/2$ we find  $y=-0.0658$ and $z=0$. In this case the cavity-mediated interactions are solely described by the term proportional to $\hat B^2$. Setting  $\phi=\pi/4$ the overlap integrals are $z=0.5854$ and $y=-0.0465$ and we expect to observe an interplay between the density-density potential and the correlated tunneling. 

\subsection{Observables}

The ground state of Hamiltonian \eqref{Eq:BH} is determined in one dimension and for a fixed number of bosons on a finite lattice $L$ with open boundaries.  The numerical program we use is based on the DMRG algorithm, we refer the readers to \cite{Chanda21} {and to Appendix \ref{app:DMRG}} for details on its implementation. In what follows we introduce and describe the observables we use in order to identify the quantum phases. 

We identify superfluidity by a non-vanishing value of the single-particle correlations. {In turn,} a phase is incompressible (insulator) when the single-particle correlations vanish and there is a finite energy gap between the ground and the excited states in the thermodynamic limit. In order to gain information on the properties of the superfluid phase, we analyze the Fourier transform of the single-particle correlations, the so-called
single-particle structure form factor, that is defined as \cite{Kraus20}:
	\begin{align}
	M_1(k)=\frac{1}{L^2} \sum_{i,j}e^{ik(i-j)}\left\langle \hat{a}_{i}^\dagger \hat{a}_j \label{M1_cavity} \right\rangle \, ,
	\end{align}
where $\langle \cdot\rangle$ denotes the expectation value over the ground state. This quantity can be experimentally revealed by means of time-of-flight measurements \cite{Greiner2002}. Depending on the value of $k$ at which $|M_1(k)|$ reveals a maximum, off-diagonal order can exhibit modulations in the phase. In the absence of the cavity, the ground-state superfluid is spatially homogeneous and characterized by a non-vanishing value of $M_1(k)$ at $k=0$. 

The superfluid (SF) phase acquires further {features} in the presence of the cavity field, depending on whether the expectation values $\langle \hat D\rangle$ and/or $\langle \hat B\rangle$ (compare Eqs.\ \eqref{Dfield} and \eqref{cavity_field}) are different from zero in the thermodynamic limit. For this purpose we identify the order parameters 
\begin{align}
&	\mathcal{O}_D=\frac{1}{L} \left|\left\langle \hat{D}\right\rangle  \right|, \label{O:D}\\
&     \label{O:B}\mathcal{O}_B= \frac{1}{2L}\left| \braket{\hat B} \right|\,,
	\end{align}
which can be measured by detecting the light at the cavity output \cite{Larson08,Landig16,Mottl11,Sierant19c}. The order parameter $\mathcal{O}_D$ signals the onset of density modulation (even-odd population imbalance), while $\mathcal{O}_B$ signals the formation of dimers along the lattice \cite{Caballero16}, namely, a so-called dimerized or bond ordered state \cite{Affleck87,Jurgensen14}. In addition to the ``normal'' SF, the emerging SF phases can be lattice Supersolid (SS) in the presence of diagonal long-range order ($\mathcal{O}_D\neq 0$ and $\mathcal{O}_B=0$); Bond SF (BSF) for homogeneous density and bond order ($\mathcal{O}_D=0$ and $\mathcal{O}_B\neq 0$), or Bond Supersolid (BSS) when both order parameters are non-vanishing. The phases and the corresponding order parameters are summarized in Table~\ref{Table}.
{The onsets of these gapless phases, i.e., SS, BSF and the BSS, occur due to {a} spontaneous breaking of {a} discrete $\mathbb{Z}_2$ lattice translational symmetry.} {Such {a} spontaneous discrete symmetry breaking is captured by the two-fold ground state degeneracy in these phases (Sec.~\ref{sec:symmetry_broken}).}

The insulating phases, {having vanishing $M_1(k)$ in the thermodynamic limit}, are classified according to the values of the population imbalance and of the bond order parameters. The Bond Insulator (BI) is characterized by $\mathcal O_B\neq 0$, the Charge-Density Wave (CDW) by  $\mathcal O_D\neq 0$, while in the Mott Insulator (MI) all order parameters here discussed vanish, see  Table~\ref{Table}.  
{Similarly to the SS, BSF, or BSS phases, {the} insulating BI and CDW phases are also $\mathbb{Z}_2$ symmetry broken phases}.
We remark that we {have} also {determined} the parity and string order parameter{s} \cite{Torre06,Rossini12} {in the resulting phases}: for the parameter regimes considered we do not find signatures of the Haldane insulator {(c.f., \cite{Chanda21})}. This is consistent with other numerical studies on globally interacting systems \cite{Sicks20}. 

	\begin{table}
		\begin{tabular}{|p{2.5cm}|p{1.5cm}||p{0.8cm}|p{0.8cm}|p{1.9cm}|} 
			\hline
			\small Phase &\small Acronyms&  \small{$\mathcal{O}_D$} & \small{$\mathcal{O}_B$}& {\rm max}\small{$|M_1(k)|$} \\ 
			\hline\hline
			\small{Mott-Insulator} & \small{MI}   & $0$    &$0$&   $0$ \\
			\small{Density Wave}&\small{CDW}&   $\ne 0$  & $0$   & $ 0$ \\
			\small{Bond Insulator}&\small{BI} & $ 0$ & $\ne 0$ & $ 0$  \\
			\small{Superfluid }&\small{SF} & $  0$ & $ 0$ & $M_1(0)$\\
			\small{Supersolid }&\small{SS}& $\ne 0$ & $0$ & $M_1(0)$ \\
			\small{Bond Superfluid}&\small{BSF} & $0$ & $\ne 0$ &$M_1(\pm \frac{\pi}{2})$ \\
			\small{Bond Supersolid}&\small{BSS}  & $\ne  0$ & $\ne 0$ &$M_1(\pm\frac{\pi}{2})$ \\
			\hline
		\end{tabular}
		\caption{Table of the quantum phases of the ground state of Eq. \eqref{Eq:BH}, of their acronyms, and of the corresponding behavior of the order parameters.}\label{Table}
		%			\center
	\end{table}

\section{Ground-state phase diagram}\label{Sec:4}

We determine the phase diagrams for fixed densities  as a function of the ratios $U_1/U$ and $t/U$ that we vary as previously specified. 
{We consider in particular the densities $\rho=1/2$ and $\rho=1$ since they}
 are commensurate with the long-range potential, {thus they can give rise to insulating phases in addition to the gapless ones.}
{In our model }the ratio $U_1/U$ controls the onset of structures that support the buildup of an intra-cavity field, while $t/U$ {determines} the strength of quantum fluctuations. We sweep the ratio $U_1/U$ from positive to negative values for different $\phi$. Depending on $\phi$ we rescale $U_1$ by the maximum between the coefficients $z^2$ and $y^2$ (i.e., $\max(z^2, y^2)$), thus giving the effective strength of the cavity-induced interaction. The ground state phase diagram is calculated by means of the DMRG algorithm with open boundary conditions. 

In the following we {present the phase diagram} for a finite system  of size $L=60$ sites and identify the transition lines when the corresponding order parameter exceeds a threshold value: The line separating the incompressible and the compressible phases is set at the threshold value $\max |M_1(k)| =0.1$. Bond and density-wave order is signaled by $\mathcal{O}_D>0.02$ and $\mathcal{O}_B>0.02$, respectively. 
We also analyze the order parameters across different transitions for different system-sizes $L \in [40, 120]$ in order to determine the nature of the phase transitions and to verify the stability of the phase diagram with the varying system-size.
We finally determine the central charge of the bond ordered gapless phases, BSF and BSS, that are due to the global interactions.

\subsection{Phase diagrams for $\phi=0$}

		\begin{figure}[t]
		\centering
		\includegraphics[width=\linewidth]{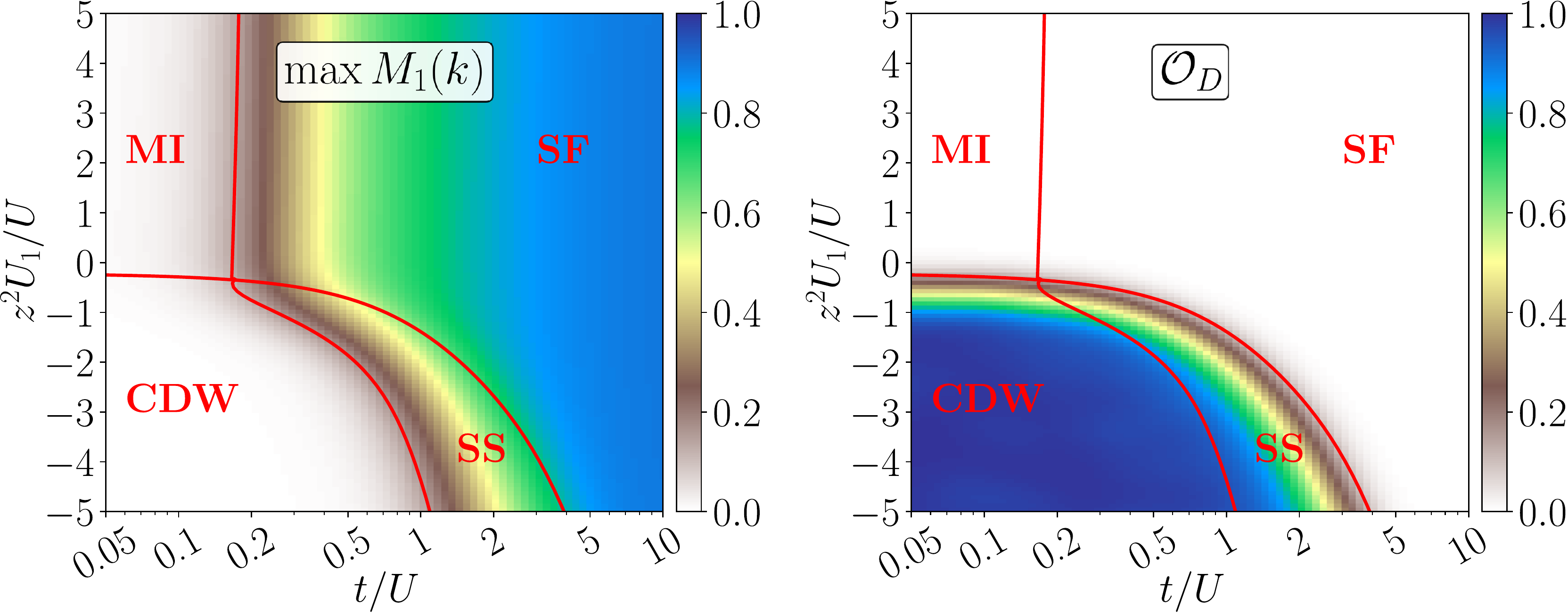}\\
		\includegraphics[width=\linewidth]{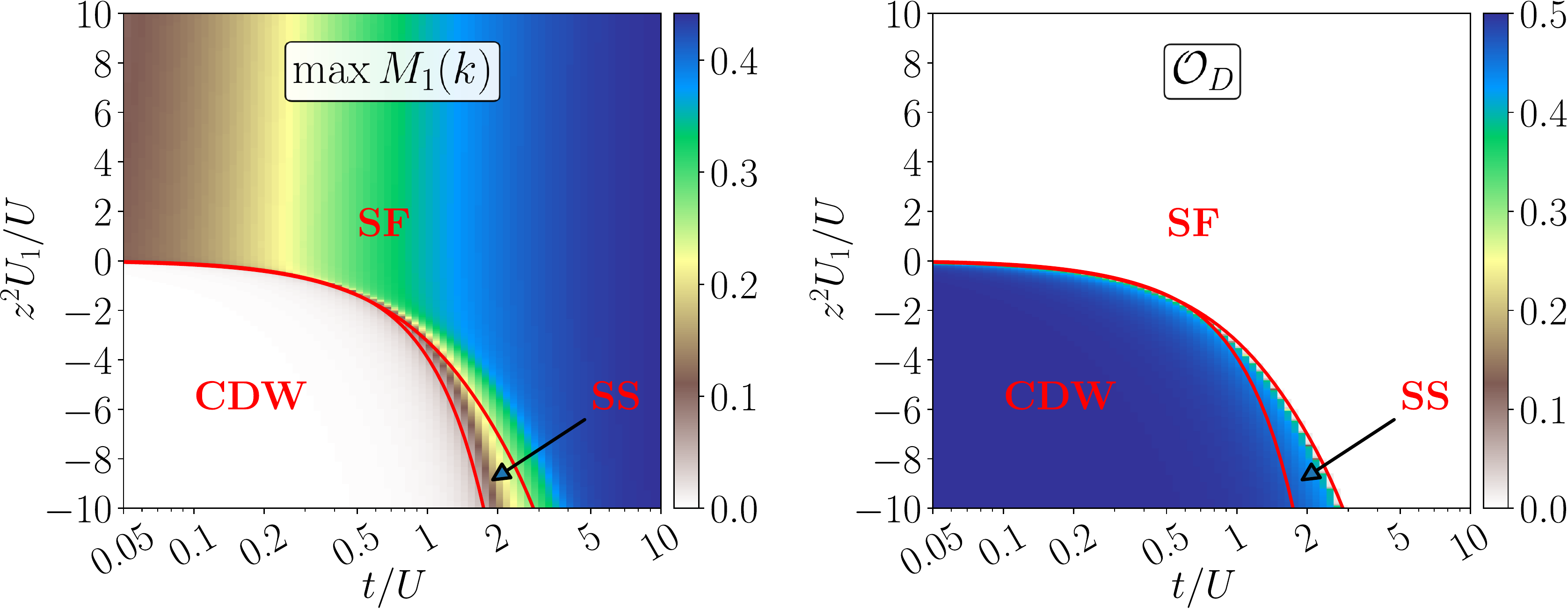}
		\caption{(Color online.) Color plots of the phase diagrams in the plane  $U_1/U$ and $t/U$ for $\phi=0$ and density $\rho=1$ (upper row) and $\rho=1/2$ (lower row). The subplots show  the maximum of $|M_1(k)|$ {(left panels)} and the density-wave order parameter {(right panels)}, the phases are identified according to Table \ref{Table}. {In the CDW phase two neighboring sites forming a unit cell have occupations $n_i = 2 \rho$, $n_{i+1} = 0$, where $\rho=1/2,1$}. Note that the interaction strength $U_1$ in the plots is scaled by $z^2$. The phase diagram are obtained using DMRG on a lattice with size $L=60$ and open boundaries. {Here,} {$t/U$ is varied from $0.01$ to $1$ in a logarithmic scale in $101$ steps, while $U_1/U$ is varied from $-20$ to $20$ in uniform steps of width $0.1$}.}
		\label{Fig:phi0}
	\end{figure}
	
For $\phi=0$ the cavity-induced interactions in the extended Bose-Hubbard Hamiltonian consist solely of global density-density interactions. These interactions are periodic, with a periodicity that is twice the lattice periodicity. The corresponding phase diagram has been extensively studied in the literature for attractive interactions, corresponding to negative values of $U_1$ \cite{Landig16,Dogra16,Niederle16,Caballero16,Flottat17,Himbert19}. In this case the cavity potential favors the formation of ordered structures which support photon scattering into the cavity. For $U_1$ positive, on the other hand, the cavity-induced potential is repulsive and the energy is minimized for uniform densities,  {at} which $\mathcal O_D$ vanishes. 

Figure~\ref{Fig:phi0} displays the maximum of $|M_1(k)|$, signaling superfluidity, and the density-wave order parameter $\mathcal{O}_D$.  For positive $U_1$  and for half-filling the ground state remains in a SF phase for the whole $t/U$ parameter range, while at unit density we find the MI-SF transition. Interestingly, the transition line slightly depends on the value of $U_1$ and in particular is  shifted to larger values of $t/U$ as $U_1$ increases: the repulsive cavity interaction tends to stabilize the incompressible phase. 

The situation is different for attractive global interactions ($U_1<0$). Here, we identify the {transition} line $U_1^{(c)}$ separating the homogeneous phase from the density wave, which is a monotonously increasing function of $t/U$. {The transition line} qualitatively agrees with the one found by means of a mean-field ansatz for a grand-canonical ensemble \cite{Himbert19}: At half-filling {it vanishes at $t/U=0$}, $U_1^{(c)}(0)=0$, while at unit density $U_1^{(c)}(0)<0$. {A direct transition between SF and the incompressible CDW is found} at half filling and for $0>U_1^{(c)}\gtrsim -U$, while for $U_1^{(c)}\lesssim -U$ a SS phase separates SF from CDW. At unit density there is no direct CDW-SF transition: The two phases are always separated either by a MI or by a SS {phase}. We also note that the area covered by the SS phase in parameter space is larger at unit density than at half-filling.

\begin{figure}
\includegraphics[width=\linewidth]{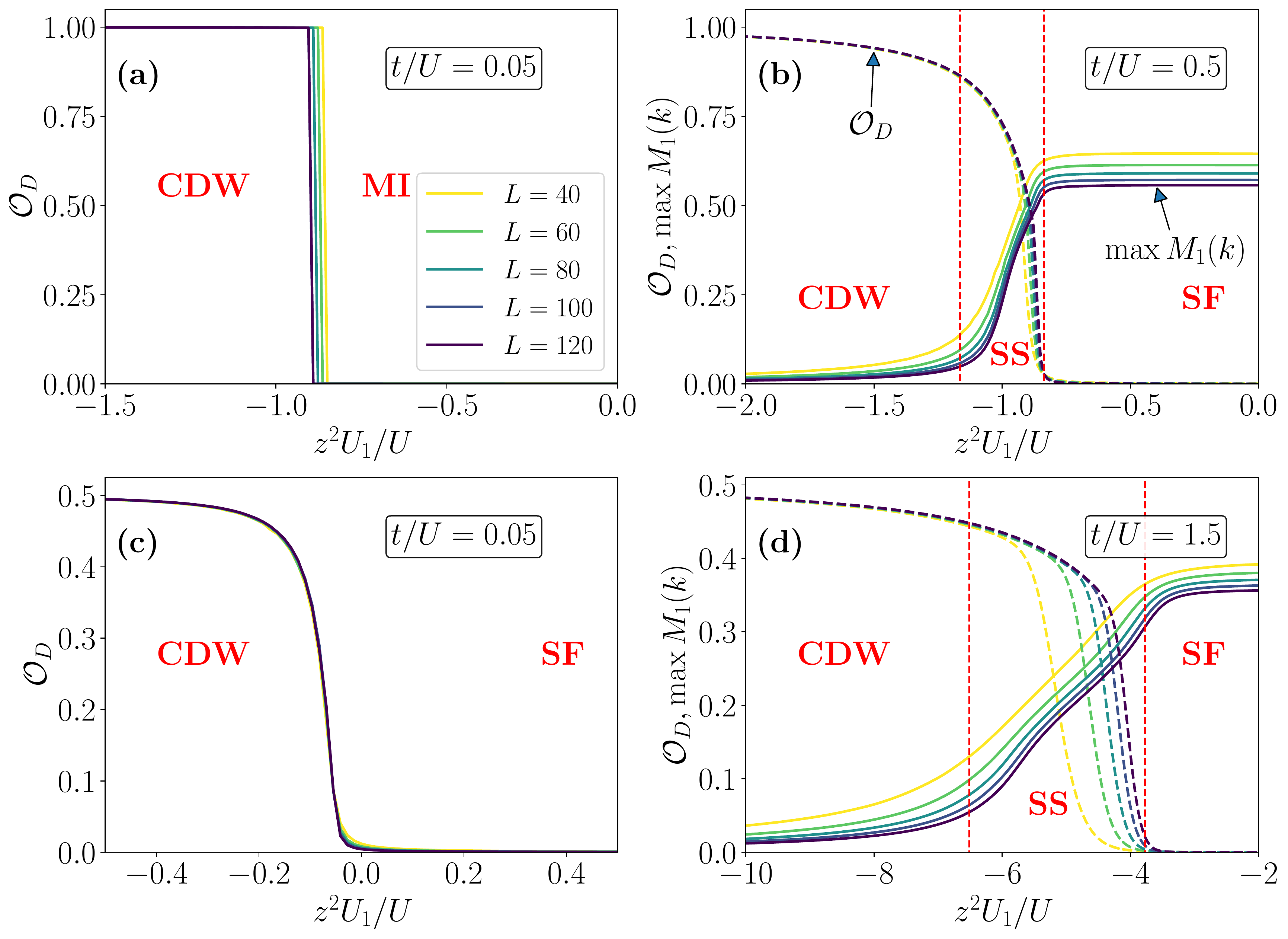}
\caption{
{(Color online.) The order parameters, $\mathcal{O}_D$ and $|\max M_1(k)|$ as a function of $U_1/U$ for $\phi=0$, fixed $t/U$ (as specified in the panels), and different system-sizes $L \in [40, 120]$ (see legend in subplot (a)).  The upper (lower) panels give the observables' behavior at  density $\rho=1$ ($\rho=1/2$) across different phase transitions. The red vertical lines indicate the value of $U_1/U$ at which we identify a phase transition, the corresponding phases are reported.
For $\rho=1$ (a) the MI-CDW transition is first-order discontinuous, while (b) the SF-SS transition is continuous and the CDW-SS transition is either continuous or a crossover.
In case of $\rho=1/2$ (c) the SF-CDW and (d) the SF-SS transitions are continuous, while the SS-CDW transition could be either continuous or a crossover.}
}
\label{fig:phi0_trans}
\end{figure}

{By inspecting the behavior of the order parameters across different phase transitions (see Fig.~\ref{fig:phi0_trans}), we deduce the nature of the transitions.
Interestingly,  for $\rho=1/2$, the transitions SF-SS and SF-CDW are now characterized by a {smooth} change in the density-wave order parameter (lower panels of Fig.~\ref{fig:phi0_trans}), signaling} {that these transitions are continuous}. The CDW-SS transition, instead, is either continuous or a crossover. {In the case of $\rho=1$ (upper panels in Fig.~\ref{fig:phi0_trans}), {a jump in the order parameter $\mathcal{O}_D$ signals a discountinuous transition between the MI and the CDW phase}.

\subsection{Phase diagrams for $\phi=\pi/2$}

		\begin{figure}[thb]
		\centering
		\includegraphics[width=\linewidth]{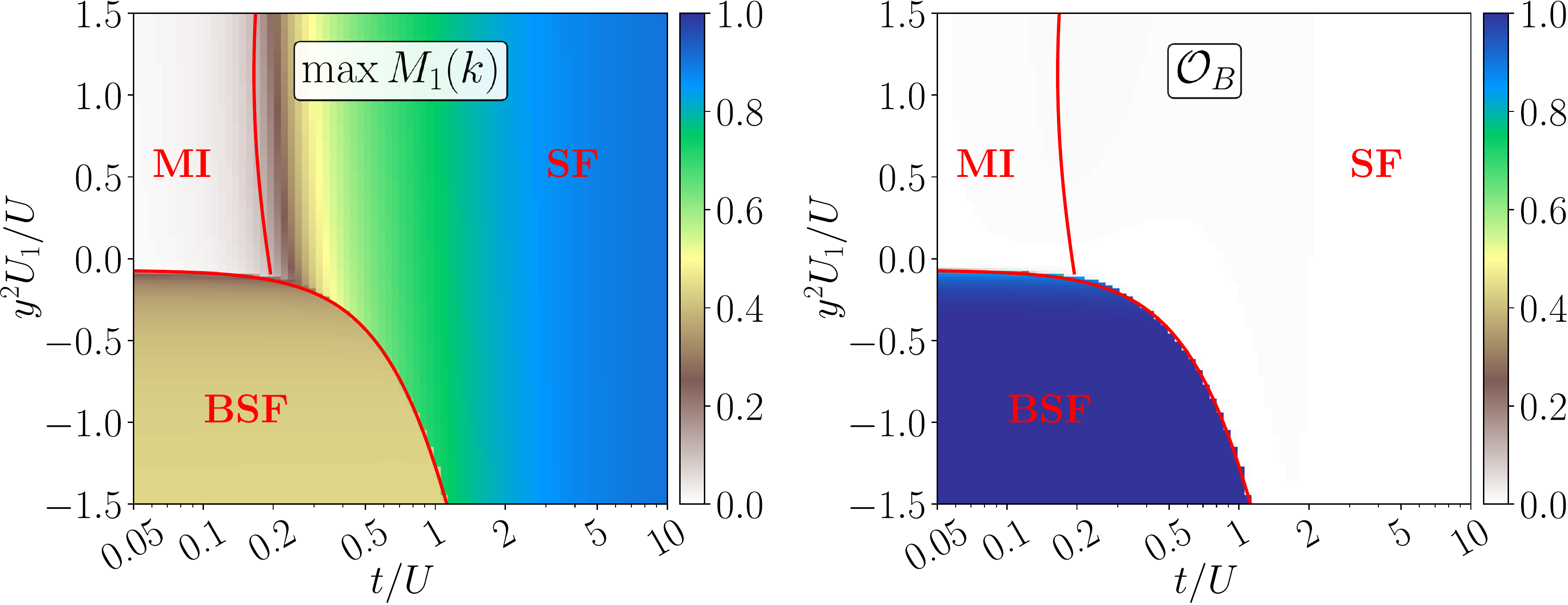}\\
		\includegraphics[width=\linewidth]{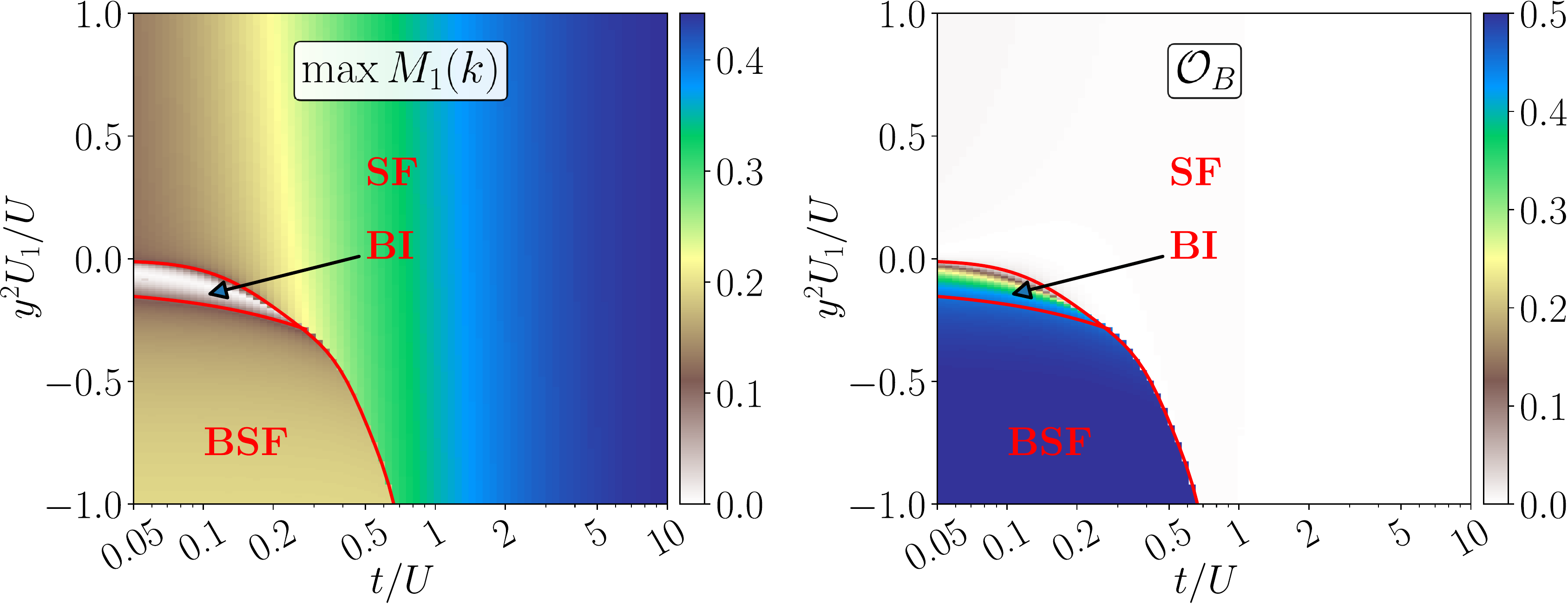}
		\caption{(Color online.) Same as Fig. \ref{Fig:phi0} but for $\phi=\pi/2$. {Here,} the right column displays  the bond order parameter, $\mathcal{O}_B$. Note that the interaction strength $U_1$ in the plots is now rescaled by $y^2$
		{and we vary $U_1/U$ in the range $[-400, 400]$ in steps of size $4$ to generate the data.}}
		\label{Fig:phi=pi2}
	\end{figure}

{We now discuss the case in which  the cavity-mediated interactions are described by a global correlated hopping term. This configuration can be realized experimentally when the atoms are tightly confined at the nodes of the cavity field. In our model, this case corresponds to the choice $\phi=\pi/2$ in the cavity standing wave, resulting in $z=0$ in Eq.~\eqref{LRBH_alphaone}.}

Figure~\ref{Fig:phi=pi2} displays the phase diagrams for density $\rho=1$ (upper row) and $\rho=1/2$ (lower row). For $U_1>0$ the ground state at half-filling is SF. At unit density, the MI-SF transition line is visibly shifted to smaller values of $t/U$ as $U_1$ increases: the size of the incompressible phase is reduced because the weight of quantum fluctuations at small $t/U$ is enhanced {due to the contribution of} the global hopping. This trend is also visible in the color plot of the maximum of $M_1(k)$ for $\rho=1/2$. 

\begin{figure}
\includegraphics[width=\linewidth]{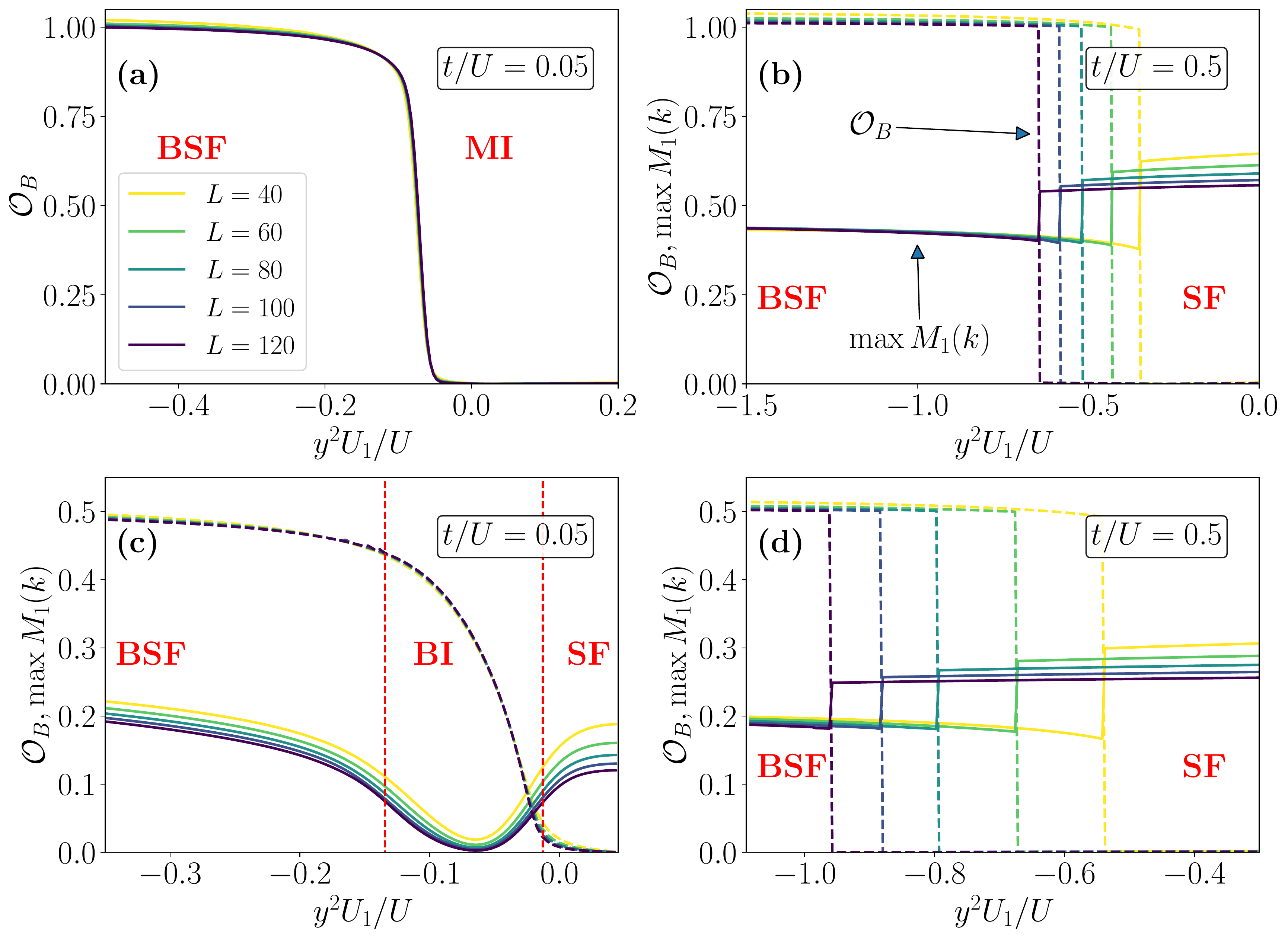}
\caption{{(Color online.) The order parameters, $\mathcal{O}_D$ and $|\max M_1(k)|$ as a function of $U_1/U$ for $\phi=\pi/2$, fixed $t/U$ (as specified in the panels), and different system-sizes $L \in [40, 120]$ (see legend in subplot (a)). The upper (lower) panels give the observables' behavior at density $\rho=1$ ($\rho=1/2$) across different phase transitions. The red vertical lines indicate the value of $U_1/U$ at which we identify a phase transition, the corresponding phases are reported.
For $\rho=1$ (a) the MI-BSF transition is continuous. Similarly, (c)  {for $\rho = 1/2$, the transitions SF-BI and BI-BSF}  are continuous. The SF-BSF transitions, on the other hand, are discontinuous for both (b) $\rho=1$ and (d) $\rho=1/2$.}
}
\label{fig:phi0p5_trans}
\end{figure}

For $U_1<0$ we observe a transition from SF to bond order. Similar to the case $\phi=0$, also here the transition shifts to larger values of $t/U$ as $|U_1|$ increases: the cavity-induced correlated hopping tends to stabilize bond order, as expected. Remarkably, at $U_1<0$ we do not find incompressible phases {for $\rho=1$}: in the considered parameter region the bond ordered phase is BSF. 
{Figure~\ref{fig:phi0p5_trans} displays the behavior of the order parameters at the transition between the BSF and the MI phase: both $M_1(k)$ and $\mathcal{O_B}$  display a continuous behavior signaling a continuous transition. On the other hand, at higher $t/U$, the transition {between} the SF {and} the BSF phase is {of} first order {kind} as visible in the discontinuity of $\max M_1(k)$ and $\mathcal{O_B}$}

The phase diagrams at $\rho=1/2$ differ from the ones at $\rho=1$ due to the appearance of an insulating phase with bond order separating the BSF from the homogeneous phase. This phase is found for sufficiently small values of $t/U$. In \cite{Chanda21} we showed that this is a topological insulator, which shares several analogies with the Su-Schrieffer–Heeger model. Inspection into the behavior of the order parameters show that the transition SF-BI and BI-BSF is continuous Fig.~\ref{fig:phi0p5_trans}), while the direct transition SF-BSF is discontinuous (c.f. \cite{Chanda21}).

\subsection{Phase diagrams for $\phi=\pi/4$}

		\begin{figure*}[t]
		\centering
		\includegraphics[width=0.767\linewidth]{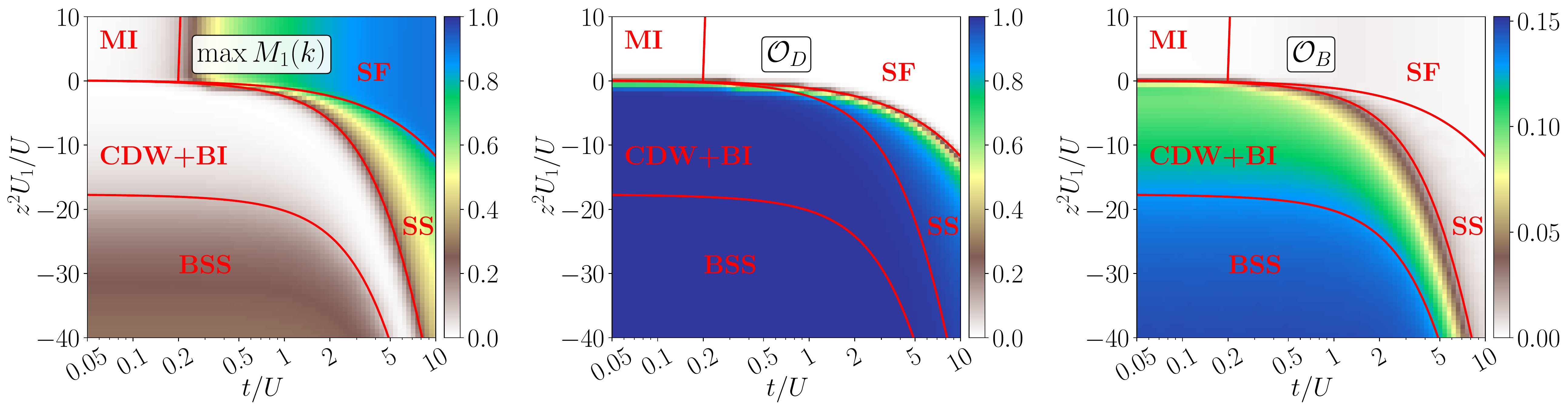}\\
		\includegraphics[width=0.767\linewidth]{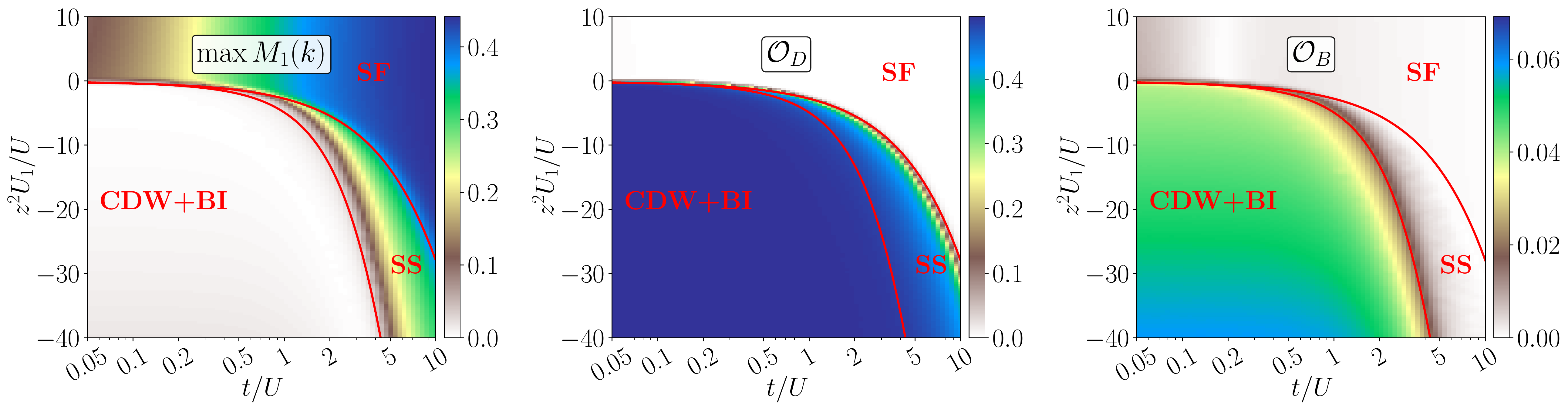}
		\caption{(Color online.) Color plots of the phase diagrams in the plane $U_1/U$ and $t/U$ at $\phi=\pi/4$ and for density $\rho=1$ (top) and $\rho=1/2$ (bottom). The subplots show the maximum of $|M_1(k)|$ (left panels), the density-wave order parameter (central panels) and the bond order parameter (right panels); the phases are labeled according to Table \ref{Table}. The phase diagrams are calculated using DMRG on a lattice with $L=60$ sites and open boundary conditions, {where we have varied $U_1/U$ from $-200$ to $200$ in steps of size $2$}. Note that the interaction strength $U_1$ is scaled by $z^2$  in all subplots.				
		}
		\label{Fig:phi=pi4-rho1}
	\end{figure*}

We now discuss the ground state phase diagrams emerging from the interplay of the density-density attractive potential and the correlated tunneling. We choose $\phi=\pi/4$ for which both $z$ and $y$ in Eq.~\eqref{LRBH_alphaone} are different from zero, and recall that $z\sim 10|y|${, see Fig. \ref{Fig1}}. Therefore, the coefficient scaling correlated tunneling is one order of magnitude smaller than the {coefficient}  scaling the cavity-induced potential term.

\begin{figure}
\includegraphics[width=\linewidth]{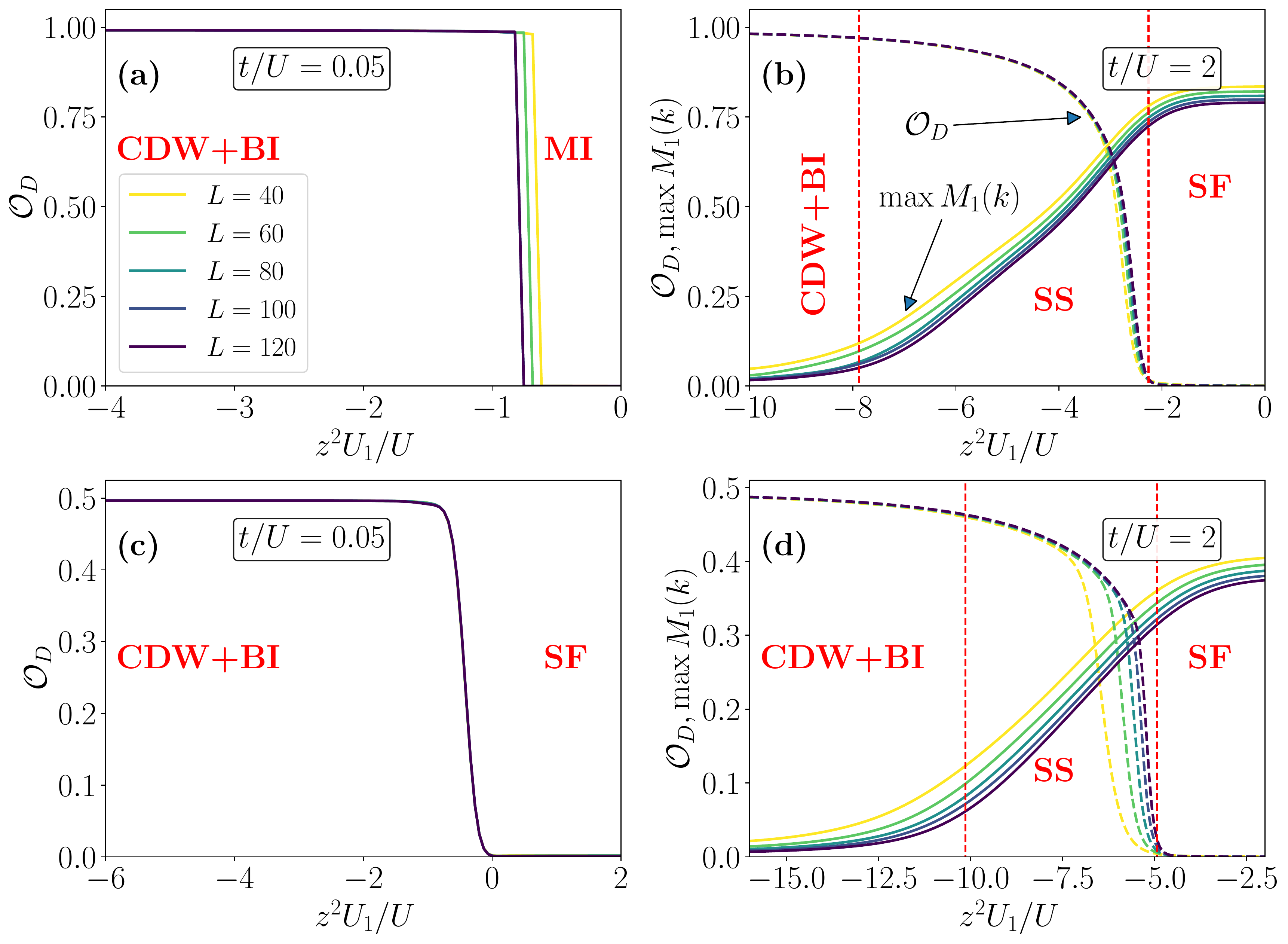}
\caption{
{(Color online.) The order parameters, $\mathcal{O}_D$ and $|\max M_1(k)|$ as a function of $U_1/U$ for $\phi=\pi/4$, fixed $t/U$ (as specified in the panels), and different system-sizes $L \in [40, 120]$ (see legend in subplot (a)).  The upper (lower) panels give the observables' behavior at  density $\rho=1$ ($\rho=1/2$) across different phase transitions. The red vertical lines indicate the value of $U_1/U$ at which we identify a phase transition, the corresponding phases are reported.
For $\rho=1$ the MI to the CDW+BI transition (a) is first-order discontinuous, while SF-SS transition (b) is continuous. The CDW+BI to the BSS (a) transition is either continuous or a crossover.
In case of $\rho=1/2$, the SF to the CDW+BI (c) and the SF-SS (d) transitions are continuous, while the {transition }SS {to} CDW+BI (d) is either continuous or a crossover.}
}
\label{fig:phi0p25_trans}
\end{figure}

Figure~\ref{Fig:phi=pi4-rho1} displays the order parameters for SF, density-wave and bond order for unit density and for half-filling. For $U_1>0$ the behavior we observe is essentially the same as for $\phi=0$.  {Instead,}
for $U_1<0$ the phase diagram becomes richer. In the first place,  at small values of $t/U$ we observe a transition from the homogeneous phases to an insulating phase with both density-wave and bond order. This transition occurs for both density $\rho=1$ and $1/2$. The new phase is an insulator of dimers with population imbalance within the dimer. We dub this phase CDW+BI since both order parameters $\mathcal O_D$ and $\mathcal O_B$ are non-zero. 
{As visible in Fig.~\ref{fig:phi0p25_trans},} the transition from the SF/SS to the CDW+BI at unit density is continuous, while the MI to the CDW+BI transition is discontinuous. For $\rho=1/2$, instead, all the transitions are continuous or crossovers.
At larger tunneling rates $t/U$ the CDW+BI phase undergoes a transition to a SS phase for both $\rho=1$ and $1/2$. The SS phase is signaled by the non-zero values of the maximum of $M_1(k)$ and by the vanishing value of the bond order parameter $\mathcal{O}_B$. The density-wave order parameter, instead, stays finite across the transition. We note that the transition CDW+BI to SS is continuous (or a crossover) for both unit density and half filling (Fig.~\ref{fig:phi0p25_trans}). The SS phase is then separated from the SF phase by a continuous transition for both $\rho=1$ and $1/2$ (analogously to the $\phi=0$ geometry).

We now discuss the phases encountered keeping $t/U$ fixed and  tuning $U_1/U$ to larger values along the negative axis. At unit density we observe a transition from CDW+BI to  a compressible phase that has both density-wave and bond order. This phase is a SS phase exhibiting dimers -- the BSS phase.
Hence, at unit density  and for low tunneling rates, large cavity-mediated interactions promote superfluidity, which is at first sight seems counterintuitive.
Interestingly, for the parameter window we have considered, we do not find a BSS phase at half filling. However, from the pattern of $M_1(k)$, we suspect that the BSS will also appear at half filling but at larger negative values of $U_1/U$.

\subsection{Discussion}
The phase diagrams for the three geometries have been also analyzed in Ref.~\cite{Caballero16} using exact diagonalization and small chains. In this work, we refrained from making a systematic comparison of our predictions with the results of \cite{Caballero16}. {In fact,} our results are qualitatively and quantitatively different in most regions of the phase diagram. We believe that the discrepancy is mainly due to the very small size considered in \cite{Caballero16}. To give {few examples}, in  \cite{Caballero16} and at half filling the authors reported insulating phases for $U_1>0$, while instead the phase we find is always SF. Other phases, such as the SFD dimers (that would here correspond to {a sort of} BSF phase) are reported in regions of the phase diagram where the DMRG predicts different ground state phases. A Gutzwiller mean-field analysis confirms the DMRG result and often finds that the SFD (or BSF) phase in those regions is a metastable, excited state. 

{Remarkably, the phase diagram in Fig. \ref{fig:phi0_trans} at $\rho=1/2$ is in qualitative agreement with the mean-field predictions for a grand-canonical ensemble \cite{Dogra16,Himbert19}, despite the fact that in the present paper it has been determined using DMRG in one dimension. There are instead qualitative differences when comparing the phase diagram at unit density. } {For $\phi=\pi/2$ an analysis based on a Gutzwiller mean-field for a canonical ensemble qualitatively reproduces the DMRG phase diagram for $\rho=1$. It does not capture, however, the BI phase at half filling, see Ref.~\cite{Chanda21}.}

\subsection{Symmetry-broken gapless phases}
\label{sec:symmetry_broken}

{We have shown that the attractive cavity potential stabilizes inhomogeneous gapless phases, the SS, the BSF, and the BSS phase. These phases have long-range diagonal order,  signaled by the non-vanishing order parameters $\mathcal{O}_D$ and/or $\mathcal{O}_B$, and off-diagonal correlations that decay algebraically with the distance. }
The long-range order in these three gapless phases manifests {itself} due to {a} spontaneous breaking of discrete $\mathbb{Z}_2$ lattice translational symmetry by doubling of the unit cell.
{A way} to verify such {a} spontaneous symmetry breaking, is by checking the ground state degeneracy for finite system-sizes. For spontaneous $\mathbb{Z}_2$ symmetry breaking, the energy gap between the ground state and the first excited state must fall to zero much faster than $\sim 1/L$ (ideally, the gap should diminish exponentially in the system-size), while the spectral gap, as measured by the gap 
between the ground state and the second excited state should vanish as $\sim 1/L$ in a gapless phase.

\begin{figure}
\centering
	\includegraphics[width=\linewidth]{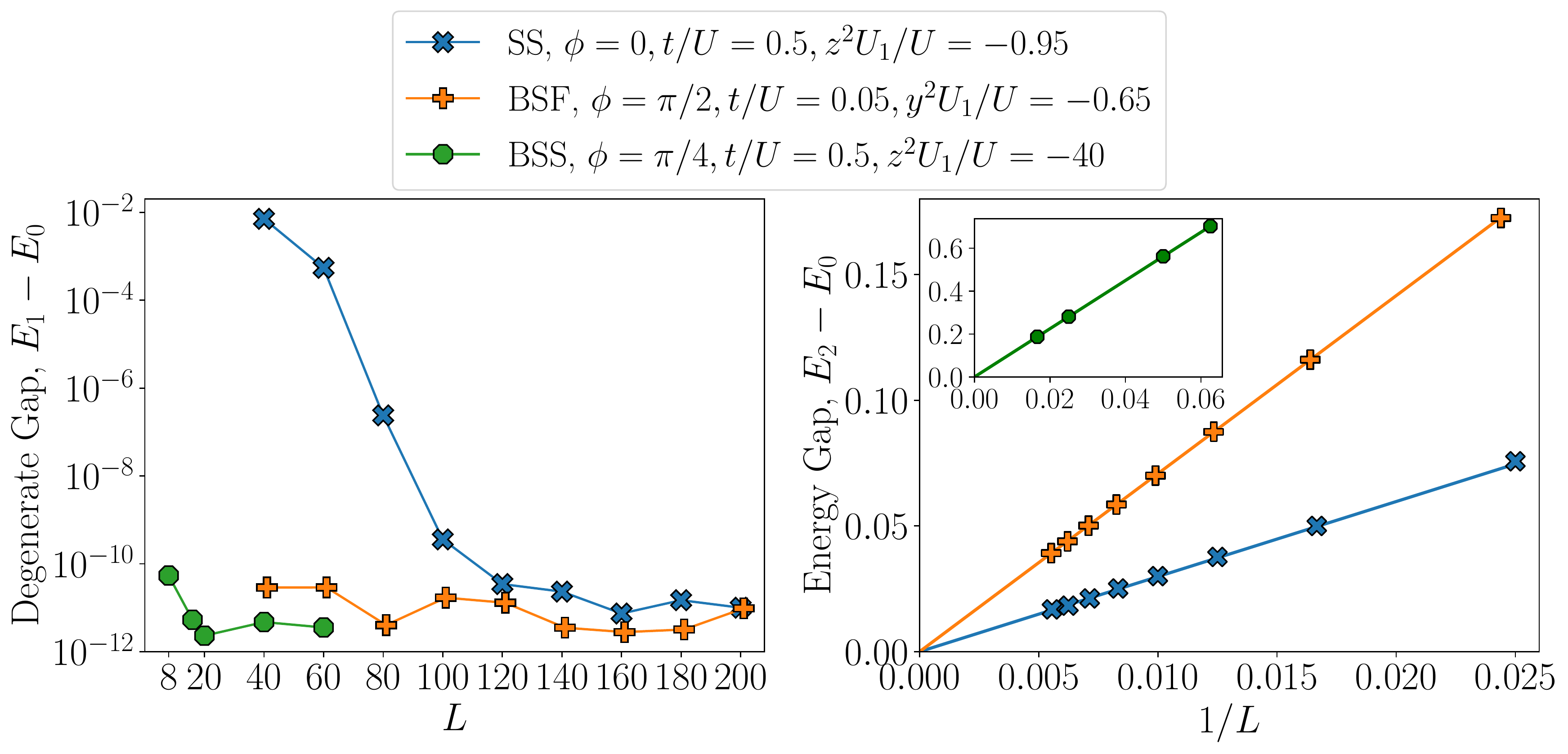}
	\caption{{(Color online.) The energy gaps in the SS, the BSF, and the BSS phase at unit filling. In the left panel, we plot the energy gap between the ground and the first excited states ($E_1-E_0$) as a function of the system-size $L$ confirming the ground state degeneracy in these phases.
The right panel shows the gap between the ground and the second excited states ($E_2-E_0$) that decays to zero in the thermodynamic limit as $\sim 1/L$ confirming the gapless character of the phases. Here we consider {an} even number of sites for the SS phase and {an} odd number of sites for the BSF phase with open boundary condition, while for BSS phase we impose periodic boundary condition with even number of sites.}}
	\label{fig:egap}
\end{figure}

 In the SS phase ($O_D \neq 0$), the order develops in the sites by $\mathbb{Z}_2$ symmetry breaking, and therefore the ground state must be two-fold degenerate for even number of sites for large enough system-sizes.
{In the left panel of Fig.~\ref{fig:egap}} {we plot the degenerate gap, i.e., the gap between the ground and the first excited states ($E_1-E_0$). {The gap}  reaches the numerical precision $(\lesssim 10^{-10})$ set by DMRG simulations as we increase the system-size.} {The right panel} {show{s} the gap between the ground and the second excited states ($E_2-E_0$)} {as a function of $1/L$. It decays to zero as $\sim 1/L$, confirming the gapless nature of the phase.}
{The number of sites shall be odd when analyzing the BSF phase ($O_B \neq 0$), because the order develops in the bonds.}  {Therefore, even number of bonds (and hence odd number of sites) are needed
to observe the ground state degeneracy. In Fig.~\ref{fig:egap} we verify the ground state degeneracy along with gapless nature of the BSF phase for odd number of sites ($L=41, 61, 81,...$).
In the BSS phase ($O_D$ and $O_B$ both are non-zero), the order develops in both sites and bonds. Therefore, we need to impose periodic boundary condition with even number of sites/bonds to observe the  degeneracy. We find that, even for system-size $L=8$ with periodic boundary condition, the  gap in the 
BSS phase  reaches the numerical precision (Fig.~\ref{fig:egap}),}  {thus we expect that it vanishes in the thermodynamic limit.}

\subsection{Scaling of the entanglement entropy in the gapless bond ordered phases}

Let us now analyze the two bond superfluid phases we find, namely the BSF for $\phi=\pi/2$ and the BSS for $\phi=\pi/4$ and at unit density. We determine in particular the scaling of the entanglement entropy with the system size.  The entanglement entropy of a lattice partition comprising the sites  $\ell=1,\ldots,j$ is defined as
\begin{equation}
\mathcal{S}_j = - \text{Tr} \left[\rho_j \ln \rho_j \right],
\end{equation}
where $j$ denotes the bond that separates the system into two parts, and $\rho_j = \text{Tr}_{j+1, j+2, ..., L} \ket{\psi} \bra{\psi}$ is the reduced density matrix obtained from the ground state $\ket{\psi}$ by tracing out the degrees of freedom of the second partition. In a gapless critical system with open boundary conditions the entanglement entropy scales with the size of the block of consecutive sites according to \cite{Callan94,Vidal03b,Calabrese04}
\begin{equation}
\mathcal{S}_j = \frac{c}{6} \ln \left[ \frac{2L}{\pi} \sin \left(\pi j/L\right) \right] +b',
\label{eq:cardy_calabrese}
\end{equation}
where $c$ is the central charge of the corresponding conformal field theory (CFT) that describes the criticality and $b'$ is a non-universal constant. In Bose-Hubbard models with short-range interaction, the gapless critical phases, e.g., SF and SS phases, obey the entropy scaling formula \eqref{eq:cardy_calabrese} with $c=1$ that corresponds to the CFT of free compactified bosons described by Tomonaga-Luttinger liquid theory~\cite{Cazalilla11}. However, the fate of such an entropy scaling in the presence of infinite-range global interactions is still an open question. This question is particularly intriguing 
when considering the BSF and BSS phases in our study, since these gapless phases are due to the infinite-range interactions.

\begin{figure}
		\includegraphics[width=\linewidth]{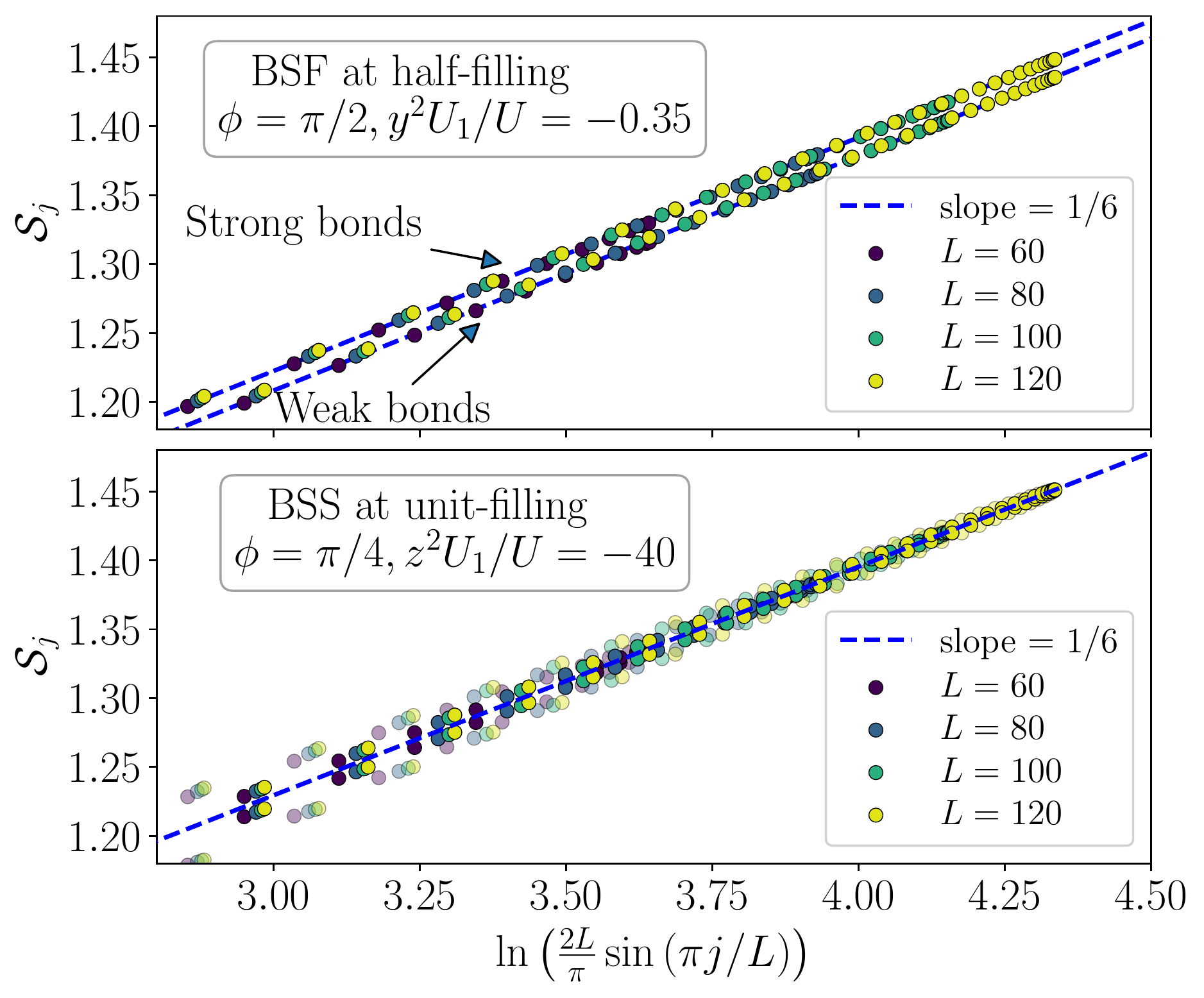}
	\caption{(Color online.) The scaling of entanglement entropy in the BSF (top panel) and the BSS (bottom panel) phases according to the formula {in Eq,\ } \eqref{eq:cardy_calabrese}. For both cases $t=0.5U$, the other parameters are given in the legends. The numerical fits yield $c = 1.00(2)$ and $c = 0.99(5)$ respectively. Note that for the BSF we get two distinct straight lines, both having a slope of $\sim 1/6$, corresponding to the entanglement entropy when the partition is cut at a strong and at a weak bond respectively. In the BSS the fluctuations of the entropy with respect to the mean slope of $1/6$ is due to the modulations of densities and bonds across the system.
}
	\label{Fig:entScaling}
\end{figure}

Figure~\ref{Fig:entScaling} displays the entanglement entropy as a function of the chord length $\ln \left[ \frac{2L}{\pi} \sin \left(\pi j/L\right) \right]$ in the BSF and the BSS phases for system sizes $L = 60, 80, 100, 120$. Interestingly, although these phases appear when the global correlated tunneling and the {potential} dominate over the short-ranged counterparts, the entropy still shows a linear growth with respect to the chord length and hence a logarithmic divergence with respect to the system size.  
Moreover, the corresponding central charges  are $c \simeq 1$, as in the cases of SF and SS phases of the  Bose-Hubbard model with short-range interactions.

\section{Conclusions}\label{Sec:5}
 
{In this work} we have presented a numerical analysis of the quantum phases of bosons in an extended Bose-Hubbard model with global density-density interactions and global correlated tunneling. The quantum phases have been studied for an one dimensional lattice and for configurations which are consistent with the setups of cavity quantum electrodynamics, where the global interactions are mediated by multiple scattered photons in the dispersive optomechanical regime \cite{Landig16,Zupancic19}. The geometry we considered permits one to {tune the relative strength of the coefficients scaling} the long-range potential and the correlated tunneling {terms in the extended Bose-Hubbard model}

{In the geometry where correlated tunneling vanishes,} the cavity-mediated potential induces density-wave order and the phase diagrams we obtain are consistent with the ones predicted  in higher dimensions~\cite{Dogra16,Niederle16,Flottat17,Himbert19}. {In the geometries where correlated tunneling cannot be discarded, we find phases with} bond order.  These phases are analogous to valence bond solids{/fluids} in spin systems~\cite{Affleck87}. We remark that bond ordered phases are also present in other systems like for instance in zig-zag optical lattices with density-dependent gauge fields~\cite{Mishra16}, in spin-1/2 dipolar Fermi gases in lattice potentials~\cite{Dio14} and in honeycomb lattices with anisotropic tunneling~\cite{Jurgensen14}. In contrast to those models here the bond order is {a self-organized phase} induced by {photon}-atom {interactions in a cavity}.

The interplay of the cavity-induced potential and correlated tunneling {gives rise to phases that can simultaneously exhibit density-wave and bond order. Remarkably, at low tunneling and unit density the cavity interactions favor a bond supersolid, where one would otherwise expect an incompressible phase.} We have analyzed the nature of {these} self-organized bond superfluid phases for exemplary parameter regimes by means of the scaling of the entanglement entropy and found that these phases have the same central charge as {the} superfluid phases {of} short-ranged Bose-Hubbard models. 

Bond and density wave order correspond to self-organized patterns which scatter coherently into the cavity mode. Correspondingly, they are associated with the onset of an intracavity field that is proportional to density-wave and bond-order parameters and can be measured at the cavity output \cite{Baumann10,Fernandez10,Sierant19c}. Superfluidity can be revealed by time-of-flight measurements \cite{Baumann10}. The low energy gap could be determined by means of a pump-probe experiment.

Bond order in these dynamics can be often understood in terms of interference between single-particle hopping and correlated tunneling \cite{Kraus20,Chanda21}. The very same interference is at the basis of the phases observed in one dimension for correlated tunneling with short range interactions, even though the resulting phases are different \cite{Biedron18,Kraus20}. In two dimensions the phases result from an interesting interplay between frustration and quantum interference \cite{Suthar20,Zhang21}. The extension of these dynamics to global interactions is non-trivial and will be subject of future works.

\acknowledgements
R.K. and G.M. acknowledge the support of the Deutsche Forschungsgemeinschaft (DFG, German Research Foundation), Project-ID 429529648 TRR 306 QuCoLiMa (``Quantum Cooperativity of Light and Matter'') and Priority Program SPP1929 GiRyd (``Giant Interactions in Rydberd Systems''). They also acknowledge the support of the German Ministry of Education and Research (BMBF) via the QuantERA projects ``QTFLAG'' and ``NAQUAS''. Project NAQUAS has received funding from the QuantERA ERA-NET Cofund in QuantumTechnologies implemented within the European Union's Horizon 2020 Programme. %The numerical DMRG computations have been possible thanks to PL-Grid Infrastructure.
T.C. and J.Z. acknowledge the support of  PL-Grid Infrastructure and the National Science Centre (Poland) under project Opus 2019/35/B/ST2/00034 (J.Z.) and Unisono 2017/25/Z/ST2/03029 (T.C.) realized within QuantERA ERA-NET QTFLAG collaboration.
{The DMRG simulations have been performed using ITensor library \cite{itensor}.}

\begin{appendix}

\section{On the extended Bose-Hubbard model of cavity quantum electrodynamics}
\label{app:BH}

The extended Bose-Hubbard model of Eq. \eqref{Eq:BH} is found by performing the Wannier expansion of the Hamiltonian $\hat{H}_{\text{eff}}=\hat{H}+\hat{H}_{\text{C}}$ as a function of the bosonic field operators $\Psi(\mathbf{r})$ with $\left[\Psi(\mathbf{r}),\Psi^\dagger(\mathbf{r}') \right] = \delta(\mathbf{r}-\mathbf{r}')$. The Hamiltonian term $\hat{H}$ consists of the kinetic energy, the potential of the optical lattice and contact interaction between the atoms:
	\begin{align}
	\hat{H}= &\int d^3\mathbf{r}\hat{\Psi}^\dagger(\mathbf{r})\left(-\frac{\hbar^2}{2m}\nabla^2 + V_{\text{trap}} \right)\hat{\Psi}(\mathbf{r}) \\
	&+ \frac{g}{2}\int d^3\mathbf{r}\hat{\Psi}^\dagger(\mathbf{r})^2\hat{\Psi}(\mathbf{r})^2  \,,
	\end{align}
while the effect of the atom-cavity coupling is described by the effective long-range Hamiltonian \cite{Fernandez10,Habibian13b}
	\begin{align}
		\hat{H}_C= \frac{U_1}{L}\left( \int d^3\mathbf{r}\hat{\Psi}^\dagger(\mathbf{r})  h(\mathbf{r})\hat{\Psi}(\mathbf{r}) \right)^2  \,, \label{LR}
	\end{align}
where $h(\mathbf{r})$ is the spatial mode function of the cavity field, with $\max |h(\mathbf{r})|=1$. Hamiltonian \eqref{LR} is obtained by eliminating the cavity field assuming that this evolves on a faster time-scale than the atomic motion. The details of the derivation of Eq. \eqref{LR} starting from the full atom-photon master equation (that accounts for cavity losses) have been reported and discussed in \cite{Larson08,Fernandez10,Habibian13,Sierant19c}, see also Ref. \cite{Schuetz13} for a systematic semiclassical treatment. We note that in Eq. \eqref{LR} we have neglected a term resulting from the dynamical Stark shift of the cavity field. We further remark that the coefficient $U_1$ is proportional to the intensity of a transverse laser field, driving the atoms, and that its sign is experimentally controlled by the sign of the detuning between the cavity and laser frequency \cite{Fernandez10,Habibian13,Ritsch13}. 

In our work we assume a trapping potential of the form 
 $V_{\text{trap}}= V_{\text{lat}}(x)+ \frac{m\omega^2}{2}\left(y^2+ z^2 \right)$ with $\omega$ the trap frequency and 
$$V_{\text{lat}}(x)=V_0\sin^2(kx).$$
The trap frequency $\omega$ is chosen so that the atoms can be assumed to be in the ground state of the transverse trapping potential. In the single-band approximation the Wannier expansion of the bosonic field reads 
	\begin{align} 
	\label{Eq:Psi}
		\Psi(\mathbf{r})=\sum_j w_j(x)\psi_0(y,z)\hat a_j,
	\end{align}	
where $w_j(x)$ are one dimensional Wannier functions centered on $j$-sites along $x$ direction, $\psi_0(y,z)$ is the ground state wave-function of the transverse harmonic trap,  and $\hat a_j$ annihilates a bosons at site $j$ and in the ground state of the harmonic trap. 

The cavity spatial mode function is here assumed to be given by its expression in the paraxial approximation,
$$h({\bf r})=\cos(kx+\phi)\,,$$
assuming that the transverse size of the atomic gas is much smaller than the mode waist. 

By using Eq. \eqref{Eq:Psi} in Hamiltonian $\hat H$ and performing the integrals as outline in Refs. \cite{Larson08,Fernandez10,Habibian13b}, one obtains the Bose-Hubbard Hamiltonian of Eq.~\eqref{Eq:BH}.
	
\section{Details about DMRG simulations}
\label{app:DMRG}	

The DMRG  algorithm \cite{White92, White93}  that we employ is based on matrix product state (MPS) ansatz \cite{Schollwoeck11, Orus14} to find the ground state and low-lying excited states of the system. We employ the global $U(1)$ symmetry  corresponding to the conservation of the total number of particles \cite{Singh10, Singh11}. For that purpose, we use ITensor C++ library \cite{itensor}
where  the MPO for the all connected long-range Hamiltonian can be constructed exactly \cite{Crosswhite08, Pirvu10} using AutoMPO class.
In our Hamiltonian \eqref{Eq:BH} (with \eqref{LRBH_alphaone}), when both $\hat{B}$ and $\hat{D}$ terms are non-zero the MPO bond dimension turns out to be $13$.
The maximum number bosons ($n_0$) per site has been truncated to 6.

We consider random entangled states, $\ket{\psi_{\text{ini}}} = \frac{1}{\sqrt{50}} \sum_{i=0}^{49} \ket{\psi^{\text{rand}}_{i}}$, where $\ket{\psi^{\text{rand}}_{i}}$ are random product states with fixed density (either $\rho=1$ or $1/2$), as our initial states for the DMRG algorithm. The maximum bond dimension of MPS has been restricted to $\chi_{\max} = 600$,
so that the discarded weights of the singular values remain below $10^{-10}$ even in the gapless phases for system-sizes upto $L \leq 120$. It is to be noted that in case of small system-sizes or the gapped phases, the final bond dimension may not reach $\chi=600$, as in our DMRG sweeps we discard any singular values having weights below $10^{-12}$.
We verify the convergence of the DMRG algorithm by checking the deviations in energy in successive DMRG sweeps. When the energy deviation in successive sweeps falls below $10^{-12}$, we conclude that the resulting MPS is the ground state of the system.	
	
To obtain low-lying excited states, as in Fig.~\ref{fig:egap}, we first shift the Hamiltonian by a weight factor multiplied with the projector of the previously found state. 
To be precise, for finding the $n^{th}$ excited state $\ket{\psi_n}$, we search for the ground state of the shifted Hamiltonian, 
\begin{equation}
\hat{H}' = \hat{H} + W \sum_{m=0}^{n-1} \ket{\psi_m}\bra{\psi_m},
\end{equation}
 where $W$ should be guessed to be sufficiently larger than $E_n - E_0$.	
	
\end{appendix}

\bibliography{Bond_cavity_0508.bbl}

\begin{thebibliography}{58}%
\makeatletter
\providecommand \@ifxundefined [1]{%
 \@ifx{#1\undefined}
}%
\providecommand \@ifnum [1]{%
 \ifnum #1\expandafter \@firstoftwo
 \else \expandafter \@secondoftwo
 \fi
}%
\providecommand \@ifx [1]{%
 \ifx #1\expandafter \@firstoftwo
 \else \expandafter \@secondoftwo
 \fi
}%
\providecommand \natexlab [1]{#1}%
\providecommand \enquote  [1]{``#1''}%
\providecommand \bibnamefont  [1]{#1}%
\providecommand \bibfnamefont [1]{#1}%
\providecommand \citenamefont [1]{#1}%
\providecommand \href@noop [0]{\@secondoftwo}%
\providecommand \href [0]{\begingroup \@sanitize@url \@href}%
\providecommand \@href[1]{\@@startlink{#1}\@@href}%
\providecommand \@@href[1]{\endgroup#1\@@endlink}%
\providecommand \@sanitize@url [0]{\catcode `\\12\catcode `\$12\catcode
  `\&12\catcode `\#12\catcode `\^12\catcode `\_12\catcode `\%12\relax}%
\providecommand \@@startlink[1]{}%
\providecommand \@@endlink[0]{}%
\providecommand \url  [0]{\begingroup\@sanitize@url \@url }%
\providecommand \@url [1]{\endgroup\@href {#1}{\urlprefix }}%
\providecommand \urlprefix  [0]{URL }%
\providecommand \Eprint [0]{\href }%
\providecommand \doibase [0]{http://dx.doi.org/}%
\providecommand \selectlanguage [0]{\@gobble}%
\providecommand \bibinfo  [0]{\@secondoftwo}%
\providecommand \bibfield  [0]{\@secondoftwo}%
\providecommand \translation [1]{[#1]}%
\providecommand \BibitemOpen [0]{}%
\providecommand \bibitemStop [0]{}%
\providecommand \bibitemNoStop [0]{.\EOS\space}%
\providecommand \EOS [0]{\spacefactor3000\relax}%
\providecommand \BibitemShut  [1]{\csname bibitem#1\endcsname}%
\let\auto@bib@innerbib\@empty
%</preamble>
\bibitem [{\citenamefont {Fisher}\ \emph {et~al.}(1989)\citenamefont {Fisher},
  \citenamefont {Weichman}, \citenamefont {Grinstein},\ and\ \citenamefont
  {Fisher}}]{Fisher1989}%
  \BibitemOpen
  \bibfield  {author} {\bibinfo {author} {\bibfnamefont {M.~P.~A.}\
  \bibnamefont {Fisher}}, \bibinfo {author} {\bibfnamefont {P.~B.}\
  \bibnamefont {Weichman}}, \bibinfo {author} {\bibfnamefont {G.}~\bibnamefont
  {Grinstein}}, \ and\ \bibinfo {author} {\bibfnamefont {D.~S.}\ \bibnamefont
  {Fisher}},\ }\href {\doibase 10.1103/PhysRevB.40.546} {\bibfield  {journal}
  {\bibinfo  {journal} {Phys. Rev. B}\ }\textbf {\bibinfo {volume} {40}},\
  \bibinfo {pages} {546} (\bibinfo {year} {1989})}\BibitemShut {NoStop}%
\bibitem [{\citenamefont {Jaksch}\ \emph {et~al.}(1998)\citenamefont {Jaksch},
  \citenamefont {Bruder}, \citenamefont {Cirac}, \citenamefont {Gardiner},\
  and\ \citenamefont {Zoller}}]{Jaksch98}%
  \BibitemOpen
  \bibfield  {author} {\bibinfo {author} {\bibfnamefont {D.}~\bibnamefont
  {Jaksch}}, \bibinfo {author} {\bibfnamefont {C.}~\bibnamefont {Bruder}},
  \bibinfo {author} {\bibfnamefont {J.~I.}\ \bibnamefont {Cirac}}, \bibinfo
  {author} {\bibfnamefont {C.~W.}\ \bibnamefont {Gardiner}}, \ and\ \bibinfo
  {author} {\bibfnamefont {P.}~\bibnamefont {Zoller}},\ }\href {\doibase
  10.1103/PhysRevLett.81.3108} {\bibfield  {journal} {\bibinfo  {journal}
  {Phys. Rev. Lett.}\ }\textbf {\bibinfo {volume} {81}},\ \bibinfo {pages}
  {3108} (\bibinfo {year} {1998})}\BibitemShut {NoStop}%
\bibitem [{\citenamefont {Greiner}\ \emph {et~al.}(2002)\citenamefont
  {Greiner}, \citenamefont {Mandel}, \citenamefont {Esslinger}, \citenamefont
  {H{\"a}nsch},\ and\ \citenamefont {Bloch}}]{Greiner2002}%
  \BibitemOpen
  \bibfield  {author} {\bibinfo {author} {\bibfnamefont {M.}~\bibnamefont
  {Greiner}}, \bibinfo {author} {\bibfnamefont {O.}~\bibnamefont {Mandel}},
  \bibinfo {author} {\bibfnamefont {T.}~\bibnamefont {Esslinger}}, \bibinfo
  {author} {\bibfnamefont {T.~W.}\ \bibnamefont {H{\"a}nsch}}, \ and\ \bibinfo
  {author} {\bibfnamefont {I.}~\bibnamefont {Bloch}},\ }\href {\doibase
  10.1038/415039a} {\bibfield  {journal} {\bibinfo  {journal} {Nature}\
  }\textbf {\bibinfo {volume} {415}},\ \bibinfo {pages} {39} (\bibinfo {year}
  {2002})}\BibitemShut {NoStop}%
\bibitem [{\citenamefont {Bloch}\ \emph {et~al.}(2008)\citenamefont {Bloch},
  \citenamefont {Dalibard},\ and\ \citenamefont {Zwerger}}]{Bloch08}%
  \BibitemOpen
  \bibfield  {author} {\bibinfo {author} {\bibfnamefont {I.}~\bibnamefont
  {Bloch}}, \bibinfo {author} {\bibfnamefont {J.}~\bibnamefont {Dalibard}}, \
  and\ \bibinfo {author} {\bibfnamefont {W.}~\bibnamefont {Zwerger}},\ }\href
  {\doibase 10.1103/RevModPhys.80.885} {\bibfield  {journal} {\bibinfo
  {journal} {Rev. Mod. Phys.}\ }\textbf {\bibinfo {volume} {80}},\ \bibinfo
  {pages} {885} (\bibinfo {year} {2008})}\BibitemShut {NoStop}%
\bibitem [{\citenamefont {Lewenstein}\ \emph {et~al.}(2012)\citenamefont
  {Lewenstein}, \citenamefont {Sanpera},\ and\ \citenamefont
  {Ahufinger}}]{Lewenstein12}%
  \BibitemOpen
  \bibfield  {author} {\bibinfo {author} {\bibfnamefont {M.}~\bibnamefont
  {Lewenstein}}, \bibinfo {author} {\bibfnamefont {A.}~\bibnamefont {Sanpera}},
  \ and\ \bibinfo {author} {\bibfnamefont {V.}~\bibnamefont {Ahufinger}},\
  }\href@noop {} {\emph {\bibinfo {title} {Ultracold Atoms in Optical Lattices:
  Simulating quantum many-body systems}}}\ (\bibinfo  {publisher} {Oxford
  University Press},\ \bibinfo {year} {2012})\BibitemShut {NoStop}%
\bibitem [{\citenamefont {Gopalakrishnan}\ \emph {et~al.}(2011)\citenamefont
  {Gopalakrishnan}, \citenamefont {Lev},\ and\ \citenamefont
  {Goldbart}}]{Gopalakrishnan11}%
  \BibitemOpen
  \bibfield  {author} {\bibinfo {author} {\bibfnamefont {S.}~\bibnamefont
  {Gopalakrishnan}}, \bibinfo {author} {\bibfnamefont {B.~L.}\ \bibnamefont
  {Lev}}, \ and\ \bibinfo {author} {\bibfnamefont {P.~M.}\ \bibnamefont
  {Goldbart}},\ }\href {\doibase 10.1103/PhysRevLett.107.277201} {\bibfield
  {journal} {\bibinfo  {journal} {Phys. Rev. Lett.}\ }\textbf {\bibinfo
  {volume} {107}},\ \bibinfo {pages} {277201} (\bibinfo {year}
  {2011})}\BibitemShut {NoStop}%
\bibitem [{\citenamefont {Periwal}\ \emph {et~al.}(2021)\citenamefont
  {Periwal}, \citenamefont {Cooper}, \citenamefont {Kunkel}, \citenamefont
  {Wienand}, \citenamefont {Davis},\ and\ \citenamefont
  {Schleier-Smith}}]{Schleier-Smith21}%
  \BibitemOpen
  \bibfield  {author} {\bibinfo {author} {\bibfnamefont {A.}~\bibnamefont
  {Periwal}}, \bibinfo {author} {\bibfnamefont {E.~S.}\ \bibnamefont {Cooper}},
  \bibinfo {author} {\bibfnamefont {P.}~\bibnamefont {Kunkel}}, \bibinfo
  {author} {\bibfnamefont {J.~F.}\ \bibnamefont {Wienand}}, \bibinfo {author}
  {\bibfnamefont {E.~J.}\ \bibnamefont {Davis}}, \ and\ \bibinfo {author}
  {\bibfnamefont {M.}~\bibnamefont {Schleier-Smith}},\ }\href {\doibase
  10.1038/s41586-021-04156-0} {\bibfield  {journal} {\bibinfo  {journal}
  {Nature}\ }\textbf {\bibinfo {volume} {600}},\ \bibinfo {pages} {630}
  (\bibinfo {year} {2021})}\BibitemShut {NoStop}%
\bibitem [{\citenamefont {Landig}\ \emph {et~al.}(2016)\citenamefont {Landig},
  \citenamefont {Hruby}, \citenamefont {Dogra}, \citenamefont {Landini},
  \citenamefont {Mottl}, \citenamefont {Donner},\ and\ \citenamefont
  {Esslinger}}]{Landig16}%
  \BibitemOpen
  \bibfield  {author} {\bibinfo {author} {\bibfnamefont {R.}~\bibnamefont
  {Landig}}, \bibinfo {author} {\bibfnamefont {L.}~\bibnamefont {Hruby}},
  \bibinfo {author} {\bibfnamefont {N.}~\bibnamefont {Dogra}}, \bibinfo
  {author} {\bibfnamefont {M.}~\bibnamefont {Landini}}, \bibinfo {author}
  {\bibfnamefont {R.}~\bibnamefont {Mottl}}, \bibinfo {author} {\bibfnamefont
  {T.}~\bibnamefont {Donner}}, \ and\ \bibinfo {author} {\bibfnamefont
  {T.}~\bibnamefont {Esslinger}},\ }\href
  {http://dx.doi.org/10.1038/nature17409} {\bibfield  {journal} {\bibinfo
  {journal} {Nature}\ }\textbf {\bibinfo {volume} {532}},\ \bibinfo {pages}
  {476 EP } (\bibinfo {year} {2016})}\BibitemShut {NoStop}%
\bibitem [{\citenamefont {Dogra}\ \emph {et~al.}(2016)\citenamefont {Dogra},
  \citenamefont {Brennecke}, \citenamefont {Huber},\ and\ \citenamefont
  {Donner}}]{Dogra16}%
  \BibitemOpen
  \bibfield  {author} {\bibinfo {author} {\bibfnamefont {N.}~\bibnamefont
  {Dogra}}, \bibinfo {author} {\bibfnamefont {F.}~\bibnamefont {Brennecke}},
  \bibinfo {author} {\bibfnamefont {S.~D.}\ \bibnamefont {Huber}}, \ and\
  \bibinfo {author} {\bibfnamefont {T.}~\bibnamefont {Donner}},\ }\href
  {\doibase 10.1103/PhysRevA.94.023632} {\bibfield  {journal} {\bibinfo
  {journal} {Phys. Rev. A}\ }\textbf {\bibinfo {volume} {94}},\ \bibinfo
  {pages} {023632} (\bibinfo {year} {2016})}\BibitemShut {NoStop}%
\bibitem [{\citenamefont {Niederle}\ \emph {et~al.}(2016)\citenamefont
  {Niederle}, \citenamefont {Morigi},\ and\ \citenamefont
  {Rieger}}]{Niederle16}%
  \BibitemOpen
  \bibfield  {author} {\bibinfo {author} {\bibfnamefont {A.~E.}\ \bibnamefont
  {Niederle}}, \bibinfo {author} {\bibfnamefont {G.}~\bibnamefont {Morigi}}, \
  and\ \bibinfo {author} {\bibfnamefont {H.}~\bibnamefont {Rieger}},\ }\href
  {\doibase 10.1103/PhysRevA.94.033607} {\bibfield  {journal} {\bibinfo
  {journal} {Phys. Rev. A}\ }\textbf {\bibinfo {volume} {94}},\ \bibinfo
  {pages} {033607} (\bibinfo {year} {2016})}\BibitemShut {NoStop}%
\bibitem [{\citenamefont {Flottat}\ \emph {et~al.}(2017)\citenamefont
  {Flottat}, \citenamefont {de~Parny}, \citenamefont {H\'ebert}, \citenamefont
  {Rousseau},\ and\ \citenamefont {Batrouni}}]{Flottat17}%
  \BibitemOpen
  \bibfield  {author} {\bibinfo {author} {\bibfnamefont {T.}~\bibnamefont
  {Flottat}}, \bibinfo {author} {\bibfnamefont {L.~d.~F.}\ \bibnamefont
  {de~Parny}}, \bibinfo {author} {\bibfnamefont {F.}~\bibnamefont {H\'ebert}},
  \bibinfo {author} {\bibfnamefont {V.~G.}\ \bibnamefont {Rousseau}}, \ and\
  \bibinfo {author} {\bibfnamefont {G.~G.}\ \bibnamefont {Batrouni}},\ }\href
  {\doibase 10.1103/PhysRevB.95.144501} {\bibfield  {journal} {\bibinfo
  {journal} {Phys. Rev. B}\ }\textbf {\bibinfo {volume} {95}},\ \bibinfo
  {pages} {144501} (\bibinfo {year} {2017})}\BibitemShut {NoStop}%
\bibitem [{\citenamefont {Himbert}\ \emph {et~al.}(2019)\citenamefont
  {Himbert}, \citenamefont {Cormick}, \citenamefont {Kraus}, \citenamefont
  {Sharma},\ and\ \citenamefont {Morigi}}]{Himbert19}%
  \BibitemOpen
  \bibfield  {author} {\bibinfo {author} {\bibfnamefont {L.}~\bibnamefont
  {Himbert}}, \bibinfo {author} {\bibfnamefont {C.}~\bibnamefont {Cormick}},
  \bibinfo {author} {\bibfnamefont {R.}~\bibnamefont {Kraus}}, \bibinfo
  {author} {\bibfnamefont {S.}~\bibnamefont {Sharma}}, \ and\ \bibinfo {author}
  {\bibfnamefont {G.}~\bibnamefont {Morigi}},\ }\href {\doibase
  10.1103/PhysRevA.99.043633} {\bibfield  {journal} {\bibinfo  {journal} {Phys.
  Rev. A}\ }\textbf {\bibinfo {volume} {99}},\ \bibinfo {pages} {043633}
  (\bibinfo {year} {2019})}\BibitemShut {NoStop}%
\bibitem [{\citenamefont {Fern\'andez-Vidal}\ \emph {et~al.}(2010)\citenamefont
  {Fern\'andez-Vidal}, \citenamefont {De~Chiara}, \citenamefont {Larson},\ and\
  \citenamefont {Morigi}}]{Fernandez10}%
  \BibitemOpen
  \bibfield  {author} {\bibinfo {author} {\bibfnamefont {S.}~\bibnamefont
  {Fern\'andez-Vidal}}, \bibinfo {author} {\bibfnamefont {G.}~\bibnamefont
  {De~Chiara}}, \bibinfo {author} {\bibfnamefont {J.}~\bibnamefont {Larson}}, \
  and\ \bibinfo {author} {\bibfnamefont {G.}~\bibnamefont {Morigi}},\ }\href
  {\doibase 10.1103/PhysRevA.81.043407} {\bibfield  {journal} {\bibinfo
  {journal} {Phys. Rev. A}\ }\textbf {\bibinfo {volume} {81}},\ \bibinfo
  {pages} {043407} (\bibinfo {year} {2010})}\BibitemShut {NoStop}%
\bibitem [{\citenamefont {Maik}\ \emph {et~al.}(2013)\citenamefont {Maik},
  \citenamefont {Hauke}, \citenamefont {Dutta}, \citenamefont {Lewenstein},\
  and\ \citenamefont {Zakrzewski}}]{Maik2013}%
  \BibitemOpen
  \bibfield  {author} {\bibinfo {author} {\bibfnamefont {M.}~\bibnamefont
  {Maik}}, \bibinfo {author} {\bibfnamefont {P.}~\bibnamefont {Hauke}},
  \bibinfo {author} {\bibfnamefont {O.}~\bibnamefont {Dutta}}, \bibinfo
  {author} {\bibfnamefont {M.}~\bibnamefont {Lewenstein}}, \ and\ \bibinfo
  {author} {\bibfnamefont {J.}~\bibnamefont {Zakrzewski}},\ }\href {\doibase
  10.1088/1367-2630/15/11/113041} {\bibfield  {journal} {\bibinfo  {journal}
  {New Journal of Physics}\ }\textbf {\bibinfo {volume} {15}},\ \bibinfo
  {pages} {113041} (\bibinfo {year} {2013})}\BibitemShut {NoStop}%
\bibitem [{\citenamefont {Dutta}\ \emph {et~al.}(2015)\citenamefont {Dutta},
  \citenamefont {Gajda}, \citenamefont {Hauke}, \citenamefont {Lewenstein},
  \citenamefont {Luehmann}, \citenamefont {Malomed}, \citenamefont
  {Sowi\'{n}ski},\ and\ \citenamefont {Zakrzewski}}]{Dutta15}%
  \BibitemOpen
  \bibfield  {author} {\bibinfo {author} {\bibfnamefont {O.}~\bibnamefont
  {Dutta}}, \bibinfo {author} {\bibfnamefont {M.}~\bibnamefont {Gajda}},
  \bibinfo {author} {\bibfnamefont {P.}~\bibnamefont {Hauke}}, \bibinfo
  {author} {\bibfnamefont {M.}~\bibnamefont {Lewenstein}}, \bibinfo {author}
  {\bibfnamefont {D.-S.}\ \bibnamefont {Luehmann}}, \bibinfo {author}
  {\bibfnamefont {B.~A.}\ \bibnamefont {Malomed}}, \bibinfo {author}
  {\bibfnamefont {T.}~\bibnamefont {Sowi\'{n}ski}}, \ and\ \bibinfo {author}
  {\bibfnamefont {J.}~\bibnamefont {Zakrzewski}},\ }\href
  {http://stacks.iop.org/0034-4885/78/i=6/a=066001} {\bibfield  {journal}
  {\bibinfo  {journal} {Rep. Prog. Phys.}\ }\textbf {\bibinfo {volume} {78}},\
  \bibinfo {pages} {066001} (\bibinfo {year} {2015})}\BibitemShut {NoStop}%
\bibitem [{\citenamefont {Elliott}\ and\ \citenamefont
  {Mekhov}(2016)}]{Elliott16}%
  \BibitemOpen
  \bibfield  {author} {\bibinfo {author} {\bibfnamefont {T.~J.}\ \bibnamefont
  {Elliott}}\ and\ \bibinfo {author} {\bibfnamefont {I.~B.}\ \bibnamefont
  {Mekhov}},\ }\href@noop {} {\bibfield  {journal} {\bibinfo  {journal}
  {Physical Review A}\ }\textbf {\bibinfo {volume} {94}},\ \bibinfo {pages}
  {013614} (\bibinfo {year} {2016})}\BibitemShut {NoStop}%
\bibitem [{\citenamefont {Caballero-Benitez}\ and\ \citenamefont
  {Mekhov}(2016)}]{Caballero16}%
  \BibitemOpen
  \bibfield  {author} {\bibinfo {author} {\bibfnamefont {S.~F.}\ \bibnamefont
  {Caballero-Benitez}}\ and\ \bibinfo {author} {\bibfnamefont {I.~B.}\
  \bibnamefont {Mekhov}},\ }\href {\doibase 10.1088/1367-2630/18/11/113010}
  {\bibfield  {journal} {\bibinfo  {journal} {New Journal of Physics}\ }\textbf
  {\bibinfo {volume} {18}},\ \bibinfo {pages} {113010} (\bibinfo {year}
  {2016})}\BibitemShut {NoStop}%
\bibitem [{\citenamefont {Chanda}\ \emph {et~al.}(2021)\citenamefont {Chanda},
  \citenamefont {Kraus}, \citenamefont {Morigi},\ and\ \citenamefont
  {Zakrzewski}}]{Chanda21}%
  \BibitemOpen
  \bibfield  {author} {\bibinfo {author} {\bibfnamefont {T.}~\bibnamefont
  {Chanda}}, \bibinfo {author} {\bibfnamefont {R.}~\bibnamefont {Kraus}},
  \bibinfo {author} {\bibfnamefont {G.}~\bibnamefont {Morigi}}, \ and\ \bibinfo
  {author} {\bibfnamefont {J.}~\bibnamefont {Zakrzewski}},\ }\href {\doibase
  10.22331/q-2021-07-13-501} {\bibfield  {journal} {\bibinfo  {journal}
  {{Quantum}}\ }\textbf {\bibinfo {volume} {5}},\ \bibinfo {pages} {501}
  (\bibinfo {year} {2021})}\BibitemShut {NoStop}%
\bibitem [{\citenamefont {Strack}\ and\ \citenamefont
  {Vollhardt}(1993)}]{Strack93}%
  \BibitemOpen
  \bibfield  {author} {\bibinfo {author} {\bibfnamefont {R.}~\bibnamefont
  {Strack}}\ and\ \bibinfo {author} {\bibfnamefont {D.}~\bibnamefont
  {Vollhardt}},\ }\href {\doibase 10.1103/PhysRevLett.70.2637} {\bibfield
  {journal} {\bibinfo  {journal} {Phys. Rev. Lett.}\ }\textbf {\bibinfo
  {volume} {70}},\ \bibinfo {pages} {2637} (\bibinfo {year}
  {1993})}\BibitemShut {NoStop}%
\bibitem [{\citenamefont {Hirsch}(1994)}]{Hirsch94}%
  \BibitemOpen
  \bibfield  {author} {\bibinfo {author} {\bibfnamefont {J.}~\bibnamefont
  {Hirsch}},\ }\href@noop {} {\bibfield  {journal} {\bibinfo  {journal}
  {Physica B: Condensed Matter}\ }\textbf {\bibinfo {volume} {199}},\ \bibinfo
  {pages} {366} (\bibinfo {year} {1994})}\BibitemShut {NoStop}%
\bibitem [{\citenamefont {Amadon}\ and\ \citenamefont
  {Hirsch}(1996)}]{Amadon96}%
  \BibitemOpen
  \bibfield  {author} {\bibinfo {author} {\bibfnamefont {J.~C.}\ \bibnamefont
  {Amadon}}\ and\ \bibinfo {author} {\bibfnamefont {J.~E.}\ \bibnamefont
  {Hirsch}},\ }\href {\doibase 10.1103/PhysRevB.54.6364} {\bibfield  {journal}
  {\bibinfo  {journal} {Phys. Rev. B}\ }\textbf {\bibinfo {volume} {54}},\
  \bibinfo {pages} {6364} (\bibinfo {year} {1996})}\BibitemShut {NoStop}%
\bibitem [{\citenamefont {Schmidt}\ \emph {et~al.}(2008)\citenamefont
  {Schmidt}, \citenamefont {Dorier}, \citenamefont {L\"auchli},\ and\
  \citenamefont {Mila}}]{Schmidt08}%
  \BibitemOpen
  \bibfield  {author} {\bibinfo {author} {\bibfnamefont {K.~P.}\ \bibnamefont
  {Schmidt}}, \bibinfo {author} {\bibfnamefont {J.}~\bibnamefont {Dorier}},
  \bibinfo {author} {\bibfnamefont {A.~M.}\ \bibnamefont {L\"auchli}}, \ and\
  \bibinfo {author} {\bibfnamefont {F.}~\bibnamefont {Mila}},\ }\href {\doibase
  10.1103/PhysRevLett.100.090401} {\bibfield  {journal} {\bibinfo  {journal}
  {Phys. Rev. Lett.}\ }\textbf {\bibinfo {volume} {100}},\ \bibinfo {pages}
  {090401} (\bibinfo {year} {2008})}\BibitemShut {NoStop}%
\bibitem [{\citenamefont {Schmidt}\ \emph {et~al.}(2006)\citenamefont
  {Schmidt}, \citenamefont {Dorier}, \citenamefont {L\"auchli},\ and\
  \citenamefont {Mila}}]{Schmidt06}%
  \BibitemOpen
  \bibfield  {author} {\bibinfo {author} {\bibfnamefont {K.~P.}\ \bibnamefont
  {Schmidt}}, \bibinfo {author} {\bibfnamefont {J.}~\bibnamefont {Dorier}},
  \bibinfo {author} {\bibfnamefont {A.}~\bibnamefont {L\"auchli}}, \ and\
  \bibinfo {author} {\bibfnamefont {F.}~\bibnamefont {Mila}},\ }\href {\doibase
  10.1103/PhysRevB.74.174508} {\bibfield  {journal} {\bibinfo  {journal} {Phys.
  Rev. B}\ }\textbf {\bibinfo {volume} {74}},\ \bibinfo {pages} {174508}
  (\bibinfo {year} {2006})}\BibitemShut {NoStop}%
\bibitem [{\citenamefont {Sowi\ifmmode~\acute{n}\else \'{n}\fi{}ski}\ \emph
  {et~al.}(2012)\citenamefont {Sowi\ifmmode~\acute{n}\else \'{n}\fi{}ski},
  \citenamefont {Dutta}, \citenamefont {Hauke}, \citenamefont {Tagliacozzo},\
  and\ \citenamefont {Lewenstein}}]{Sowinski12}%
  \BibitemOpen
  \bibfield  {author} {\bibinfo {author} {\bibfnamefont {T.}~\bibnamefont
  {Sowi\ifmmode~\acute{n}\else \'{n}\fi{}ski}}, \bibinfo {author}
  {\bibfnamefont {O.}~\bibnamefont {Dutta}}, \bibinfo {author} {\bibfnamefont
  {P.}~\bibnamefont {Hauke}}, \bibinfo {author} {\bibfnamefont
  {L.}~\bibnamefont {Tagliacozzo}}, \ and\ \bibinfo {author} {\bibfnamefont
  {M.}~\bibnamefont {Lewenstein}},\ }\href {\doibase
  10.1103/PhysRevLett.108.115301} {\bibfield  {journal} {\bibinfo  {journal}
  {Phys. Rev. Lett.}\ }\textbf {\bibinfo {volume} {108}},\ \bibinfo {pages}
  {115301} (\bibinfo {year} {2012})}\BibitemShut {NoStop}%
\bibitem [{\citenamefont {Biedro\'{n}}\ \emph {et~al.}(2018)\citenamefont
  {Biedro\'{n}}, \citenamefont {\L{}\k{a}cki},\ and\ \citenamefont
  {Zakrzewski}}]{Biedron18}%
  \BibitemOpen
  \bibfield  {author} {\bibinfo {author} {\bibfnamefont {K.}~\bibnamefont
  {Biedro\'{n}}}, \bibinfo {author} {\bibfnamefont {M.}~\bibnamefont
  {\L{}\k{a}cki}}, \ and\ \bibinfo {author} {\bibfnamefont {J.}~\bibnamefont
  {Zakrzewski}},\ }\href {\doibase 10.1103/PhysRevB.97.245102} {\bibfield
  {journal} {\bibinfo  {journal} {Phys. Rev. B}\ }\textbf {\bibinfo {volume}
  {97}},\ \bibinfo {pages} {245102} (\bibinfo {year} {2018})}\BibitemShut
  {NoStop}%
\bibitem [{\citenamefont {Kraus}\ \emph {et~al.}(2020)\citenamefont {Kraus},
  \citenamefont {Biedro\ifmmode~\acute{n}\else \'{n}\fi{}}, \citenamefont
  {Zakrzewski},\ and\ \citenamefont {Morigi}}]{Kraus20}%
  \BibitemOpen
  \bibfield  {author} {\bibinfo {author} {\bibfnamefont {R.}~\bibnamefont
  {Kraus}}, \bibinfo {author} {\bibfnamefont {K.}~\bibnamefont
  {Biedro\ifmmode~\acute{n}\else \'{n}\fi{}}}, \bibinfo {author} {\bibfnamefont
  {J.}~\bibnamefont {Zakrzewski}}, \ and\ \bibinfo {author} {\bibfnamefont
  {G.}~\bibnamefont {Morigi}},\ }\href {\doibase 10.1103/PhysRevB.101.174505}
  {\bibfield  {journal} {\bibinfo  {journal} {Phys. Rev. B}\ }\textbf {\bibinfo
  {volume} {101}},\ \bibinfo {pages} {174505} (\bibinfo {year}
  {2020})}\BibitemShut {NoStop}%
\bibitem [{\citenamefont {Suthar}\ \emph {et~al.}(2020)\citenamefont {Suthar},
  \citenamefont {Kraus}, \citenamefont {Sable}, \citenamefont {Angom},
  \citenamefont {Morigi},\ and\ \citenamefont {Zakrzewski}}]{Suthar20}%
  \BibitemOpen
  \bibfield  {author} {\bibinfo {author} {\bibfnamefont {K.}~\bibnamefont
  {Suthar}}, \bibinfo {author} {\bibfnamefont {R.}~\bibnamefont {Kraus}},
  \bibinfo {author} {\bibfnamefont {H.}~\bibnamefont {Sable}}, \bibinfo
  {author} {\bibfnamefont {D.}~\bibnamefont {Angom}}, \bibinfo {author}
  {\bibfnamefont {G.}~\bibnamefont {Morigi}}, \ and\ \bibinfo {author}
  {\bibfnamefont {J.}~\bibnamefont {Zakrzewski}},\ }\href {\doibase
  10.1103/PhysRevB.102.214503} {\bibfield  {journal} {\bibinfo  {journal}
  {Phys. Rev. B}\ }\textbf {\bibinfo {volume} {102}},\ \bibinfo {pages}
  {214503} (\bibinfo {year} {2020})}\BibitemShut {NoStop}%
\bibitem [{\citenamefont {White}(1992)}]{White92}%
  \BibitemOpen
  \bibfield  {author} {\bibinfo {author} {\bibfnamefont {S.~R.}\ \bibnamefont
  {White}},\ }\href {\doibase 10.1103/PhysRevLett.69.2863} {\bibfield
  {journal} {\bibinfo  {journal} {Phys. Rev. Lett.}\ }\textbf {\bibinfo
  {volume} {69}},\ \bibinfo {pages} {2863} (\bibinfo {year}
  {1992})}\BibitemShut {NoStop}%
\bibitem [{\citenamefont {White}(1993)}]{White93}%
  \BibitemOpen
  \bibfield  {author} {\bibinfo {author} {\bibfnamefont {S.~R.}\ \bibnamefont
  {White}},\ }\href {\doibase 10.1103/PhysRevB.48.10345} {\bibfield  {journal}
  {\bibinfo  {journal} {Phys. Rev. B}\ }\textbf {\bibinfo {volume} {48}},\
  \bibinfo {pages} {10345} (\bibinfo {year} {1993})}\BibitemShut {NoStop}%
\bibitem [{\citenamefont {Schollwöck}(2011)}]{Schollwoeck11}%
  \BibitemOpen
  \bibfield  {author} {\bibinfo {author} {\bibfnamefont {U.}~\bibnamefont
  {Schollwöck}},\ }\href {\doibase https://doi.org/10.1016/j.aop.2010.09.012}
  {\bibfield  {journal} {\bibinfo  {journal} {Annals of Physics}\ }\textbf
  {\bibinfo {volume} {326}},\ \bibinfo {pages} {96} (\bibinfo {year}
  {2011})}\BibitemShut {NoStop}%
\bibitem [{\citenamefont {Or{\'u}s}(2014)}]{Orus14}%
  \BibitemOpen
  \bibfield  {author} {\bibinfo {author} {\bibfnamefont {R.}~\bibnamefont
  {Or{\'u}s}},\ }\href {\doibase https://doi.org/10.1016/j.aop.2014.06.013}
  {\bibfield  {journal} {\bibinfo  {journal} {Annals of Physics}\ }\textbf
  {\bibinfo {volume} {349}},\ \bibinfo {pages} {117 } (\bibinfo {year}
  {2014})}\BibitemShut {NoStop}%
\bibitem [{\citenamefont {Chanda}\ \emph {et~al.}(2020)\citenamefont {Chanda},
  \citenamefont {Zakrzewski}, \citenamefont {Lewenstein},\ and\ \citenamefont
  {Tagliacozzo}}]{Chanda20}%
  \BibitemOpen
  \bibfield  {author} {\bibinfo {author} {\bibfnamefont {T.}~\bibnamefont
  {Chanda}}, \bibinfo {author} {\bibfnamefont {J.}~\bibnamefont {Zakrzewski}},
  \bibinfo {author} {\bibfnamefont {M.}~\bibnamefont {Lewenstein}}, \ and\
  \bibinfo {author} {\bibfnamefont {L.}~\bibnamefont {Tagliacozzo}},\ }\href
  {\doibase 10.1103/PhysRevLett.124.180602} {\bibfield  {journal} {\bibinfo
  {journal} {Phys. Rev. Lett.}\ }\textbf {\bibinfo {volume} {124}},\ \bibinfo
  {pages} {180602} (\bibinfo {year} {2020})}\BibitemShut {NoStop}%
\bibitem [{\citenamefont {Habibian}\ \emph
  {et~al.}(2013{\natexlab{a}})\citenamefont {Habibian}, \citenamefont {Winter},
  \citenamefont {Paganelli}, \citenamefont {Rieger},\ and\ \citenamefont
  {Morigi}}]{Habibian13}%
  \BibitemOpen
  \bibfield  {author} {\bibinfo {author} {\bibfnamefont {H.}~\bibnamefont
  {Habibian}}, \bibinfo {author} {\bibfnamefont {A.}~\bibnamefont {Winter}},
  \bibinfo {author} {\bibfnamefont {S.}~\bibnamefont {Paganelli}}, \bibinfo
  {author} {\bibfnamefont {H.}~\bibnamefont {Rieger}}, \ and\ \bibinfo {author}
  {\bibfnamefont {G.}~\bibnamefont {Morigi}},\ }\href {\doibase
  10.1103/PhysRevLett.110.075304} {\bibfield  {journal} {\bibinfo  {journal}
  {Phys. Rev. Lett.}\ }\textbf {\bibinfo {volume} {110}},\ \bibinfo {pages}
  {075304} (\bibinfo {year} {2013}{\natexlab{a}})}\BibitemShut {NoStop}%
\bibitem [{\citenamefont {Habibian}\ \emph
  {et~al.}(2013{\natexlab{b}})\citenamefont {Habibian}, \citenamefont {Winter},
  \citenamefont {Paganelli}, \citenamefont {Rieger},\ and\ \citenamefont
  {Morigi}}]{Habibian13b}%
  \BibitemOpen
  \bibfield  {author} {\bibinfo {author} {\bibfnamefont {H.}~\bibnamefont
  {Habibian}}, \bibinfo {author} {\bibfnamefont {A.}~\bibnamefont {Winter}},
  \bibinfo {author} {\bibfnamefont {S.}~\bibnamefont {Paganelli}}, \bibinfo
  {author} {\bibfnamefont {H.}~\bibnamefont {Rieger}}, \ and\ \bibinfo {author}
  {\bibfnamefont {G.}~\bibnamefont {Morigi}},\ }\href {\doibase
  10.1103/PhysRevA.88.043618} {\bibfield  {journal} {\bibinfo  {journal} {Phys.
  Rev. A}\ }\textbf {\bibinfo {volume} {88}},\ \bibinfo {pages} {043618}
  (\bibinfo {year} {2013}{\natexlab{b}})}\BibitemShut {NoStop}%
\bibitem [{\citenamefont {Baumann}\ \emph {et~al.}(2010)\citenamefont
  {Baumann}, \citenamefont {Guerlin}, \citenamefont {Brennecke},\ and\
  \citenamefont {Esslinger}}]{Baumann10}%
  \BibitemOpen
  \bibfield  {author} {\bibinfo {author} {\bibfnamefont {K.}~\bibnamefont
  {Baumann}}, \bibinfo {author} {\bibfnamefont {C.}~\bibnamefont {Guerlin}},
  \bibinfo {author} {\bibfnamefont {F.}~\bibnamefont {Brennecke}}, \ and\
  \bibinfo {author} {\bibfnamefont {T.}~\bibnamefont {Esslinger}},\ }\href
  {http://dx.doi.org/10.1038/nature09009} {\bibfield  {journal} {\bibinfo
  {journal} {Nature}\ }\textbf {\bibinfo {volume} {464}},\ \bibinfo {pages}
  {1301 EP } (\bibinfo {year} {2010})},\ \bibinfo {note} {article}\BibitemShut
  {NoStop}%
\bibitem [{\citenamefont {Larson}\ \emph {et~al.}(2008)\citenamefont {Larson},
  \citenamefont {Damski}, \citenamefont {Morigi},\ and\ \citenamefont
  {Lewenstein}}]{Larson08}%
  \BibitemOpen
  \bibfield  {author} {\bibinfo {author} {\bibfnamefont {J.}~\bibnamefont
  {Larson}}, \bibinfo {author} {\bibfnamefont {B.}~\bibnamefont {Damski}},
  \bibinfo {author} {\bibfnamefont {G.}~\bibnamefont {Morigi}}, \ and\ \bibinfo
  {author} {\bibfnamefont {M.}~\bibnamefont {Lewenstein}},\ }\href {\doibase
  10.1103/PhysRevLett.100.050401} {\bibfield  {journal} {\bibinfo  {journal}
  {Phys. Rev. Lett.}\ }\textbf {\bibinfo {volume} {100}},\ \bibinfo {pages}
  {050401} (\bibinfo {year} {2008})}\BibitemShut {NoStop}%
\bibitem [{\citenamefont {Baumann}\ \emph {et~al.}(2011)\citenamefont
  {Baumann}, \citenamefont {Mottl}, \citenamefont {Brennecke},\ and\
  \citenamefont {Esslinger}}]{Mottl11}%
  \BibitemOpen
  \bibfield  {author} {\bibinfo {author} {\bibfnamefont {K.}~\bibnamefont
  {Baumann}}, \bibinfo {author} {\bibfnamefont {R.}~\bibnamefont {Mottl}},
  \bibinfo {author} {\bibfnamefont {F.}~\bibnamefont {Brennecke}}, \ and\
  \bibinfo {author} {\bibfnamefont {T.}~\bibnamefont {Esslinger}},\ }\href
  {\doibase 10.1103/PhysRevLett.107.140402} {\bibfield  {journal} {\bibinfo
  {journal} {Phys. Rev. Lett.}\ }\textbf {\bibinfo {volume} {107}},\ \bibinfo
  {pages} {140402} (\bibinfo {year} {2011})}\BibitemShut {NoStop}%
\bibitem [{\citenamefont {Sierant}\ \emph {et~al.}(2019)\citenamefont
  {Sierant}, \citenamefont {Biedro\'{n}}, \citenamefont {Morigi},\ and\
  \citenamefont {Zakrzewski}}]{Sierant19c}%
  \BibitemOpen
  \bibfield  {author} {\bibinfo {author} {\bibfnamefont {P.}~\bibnamefont
  {Sierant}}, \bibinfo {author} {\bibfnamefont {K.}~\bibnamefont
  {Biedro\'{n}}}, \bibinfo {author} {\bibfnamefont {G.}~\bibnamefont {Morigi}},
  \ and\ \bibinfo {author} {\bibfnamefont {J.}~\bibnamefont {Zakrzewski}},\
  }\href {\doibase 10.21468/SciPostPhys.7.1.008} {\bibfield  {journal}
  {\bibinfo  {journal} {SciPost Phys.}\ }\textbf {\bibinfo {volume} {7}},\
  \bibinfo {pages} {8} (\bibinfo {year} {2019})}\BibitemShut {NoStop}%
\bibitem [{\citenamefont {Affleck}\ \emph {et~al.}(1987)\citenamefont
  {Affleck}, \citenamefont {Kennedy}, \citenamefont {Lieb},\ and\ \citenamefont
  {Tasaki}}]{Affleck87}%
  \BibitemOpen
  \bibfield  {author} {\bibinfo {author} {\bibfnamefont {I.}~\bibnamefont
  {Affleck}}, \bibinfo {author} {\bibfnamefont {T.}~\bibnamefont {Kennedy}},
  \bibinfo {author} {\bibfnamefont {E.~H.}\ \bibnamefont {Lieb}}, \ and\
  \bibinfo {author} {\bibfnamefont {H.}~\bibnamefont {Tasaki}},\ }\href
  {\doibase 10.1103/PhysRevLett.59.799} {\bibfield  {journal} {\bibinfo
  {journal} {Phys. Rev. Lett.}\ }\textbf {\bibinfo {volume} {59}},\ \bibinfo
  {pages} {799} (\bibinfo {year} {1987})}\BibitemShut {NoStop}%
\bibitem [{\citenamefont {Jürgensen}\ and\ \citenamefont
  {Lühmann}(2014)}]{Jurgensen14}%
  \BibitemOpen
  \bibfield  {author} {\bibinfo {author} {\bibfnamefont {O.}~\bibnamefont
  {Jürgensen}}\ and\ \bibinfo {author} {\bibfnamefont {D.-S.}\ \bibnamefont
  {Lühmann}},\ }\href {\doibase 10.1088/1367-2630/16/9/093023} {\bibfield
  {journal} {\bibinfo  {journal} {New J. Phys.}\ }\textbf {\bibinfo {volume}
  {16}},\ \bibinfo {pages} {093023} (\bibinfo {year} {2014})}\BibitemShut
  {NoStop}%
\bibitem [{\citenamefont {Dalla~Torre}\ \emph {et~al.}(2006)\citenamefont
  {Dalla~Torre}, \citenamefont {Berg},\ and\ \citenamefont {Altman}}]{Torre06}%
  \BibitemOpen
  \bibfield  {author} {\bibinfo {author} {\bibfnamefont {E.~G.}\ \bibnamefont
  {Dalla~Torre}}, \bibinfo {author} {\bibfnamefont {E.}~\bibnamefont {Berg}}, \
  and\ \bibinfo {author} {\bibfnamefont {E.}~\bibnamefont {Altman}},\ }\href
  {\doibase 10.1103/PhysRevLett.97.260401} {\bibfield  {journal} {\bibinfo
  {journal} {Phys. Rev. Lett.}\ }\textbf {\bibinfo {volume} {97}},\ \bibinfo
  {pages} {260401} (\bibinfo {year} {2006})}\BibitemShut {NoStop}%
\bibitem [{\citenamefont {Rossini}\ and\ \citenamefont
  {Fazio}(2012)}]{Rossini12}%
  \BibitemOpen
  \bibfield  {author} {\bibinfo {author} {\bibfnamefont {D.}~\bibnamefont
  {Rossini}}\ and\ \bibinfo {author} {\bibfnamefont {R.}~\bibnamefont
  {Fazio}},\ }\href {\doibase 10.1088/1367-2630/14/6/065012} {\bibfield
  {journal} {\bibinfo  {journal} {New Journal of Physics}\ }\textbf {\bibinfo
  {volume} {14}},\ \bibinfo {pages} {065012} (\bibinfo {year}
  {2012})}\BibitemShut {NoStop}%
\bibitem [{\citenamefont {{Sicks, Johannes}}\ and\ \citenamefont {{Rieger,
  Heiko}}(2020)}]{Sicks20}%
  \BibitemOpen
  \bibfield  {author} {\bibinfo {author} {\bibnamefont {{Sicks, Johannes}}}\
  and\ \bibinfo {author} {\bibnamefont {{Rieger, Heiko}}},\ }\href {\doibase
  10.1140/epjb/e2020-10109-3} {\bibfield  {journal} {\bibinfo  {journal} {Eur.
  Phys. J. B}\ }\textbf {\bibinfo {volume} {93}},\ \bibinfo {pages} {104}
  (\bibinfo {year} {2020})}\BibitemShut {NoStop}%
\bibitem [{\citenamefont {Callan}\ and\ \citenamefont
  {Wilczek}(1994)}]{Callan94}%
  \BibitemOpen
  \bibfield  {author} {\bibinfo {author} {\bibfnamefont {C.}~\bibnamefont
  {Callan}}\ and\ \bibinfo {author} {\bibfnamefont {F.}~\bibnamefont
  {Wilczek}},\ }\href {\doibase 10.1016/0370-2693(94)91007-3} {\bibfield
  {journal} {\bibinfo  {journal} {Physics Letters B}\ }\textbf {\bibinfo
  {volume} {333}},\ \bibinfo {pages} {55} (\bibinfo {year} {1994})}\BibitemShut
  {NoStop}%
\bibitem [{\citenamefont {Vidal}\ \emph {et~al.}(2003)\citenamefont {Vidal},
  \citenamefont {Latorre}, \citenamefont {Rico},\ and\ \citenamefont
  {Kitaev}}]{Vidal03b}%
  \BibitemOpen
  \bibfield  {author} {\bibinfo {author} {\bibfnamefont {G.}~\bibnamefont
  {Vidal}}, \bibinfo {author} {\bibfnamefont {J.~I.}\ \bibnamefont {Latorre}},
  \bibinfo {author} {\bibfnamefont {E.}~\bibnamefont {Rico}}, \ and\ \bibinfo
  {author} {\bibfnamefont {A.}~\bibnamefont {Kitaev}},\ }\href {\doibase
  10.1103/PhysRevLett.90.227902} {\bibfield  {journal} {\bibinfo  {journal}
  {Phys. Rev. Lett.}\ }\textbf {\bibinfo {volume} {90}},\ \bibinfo {pages}
  {227902} (\bibinfo {year} {2003})}\BibitemShut {NoStop}%
\bibitem [{\citenamefont {Calabrese}\ and\ \citenamefont
  {Cardy}(2004)}]{Calabrese04}%
  \BibitemOpen
  \bibfield  {author} {\bibinfo {author} {\bibfnamefont {P.}~\bibnamefont
  {Calabrese}}\ and\ \bibinfo {author} {\bibfnamefont {J.}~\bibnamefont
  {Cardy}},\ }\href {\doibase 10.1088/1742-5468/2004/06/p06002} {\bibfield
  {journal} {\bibinfo  {journal} {Journal of Statistical Mechanics: Theory and
  Experiment}\ }\textbf {\bibinfo {volume} {2004}},\ \bibinfo {pages} {P06002}
  (\bibinfo {year} {2004})}\BibitemShut {NoStop}%
\bibitem [{\citenamefont {Cazalilla}\ \emph {et~al.}(2011)\citenamefont
  {Cazalilla}, \citenamefont {Citro}, \citenamefont {Giamarchi}, \citenamefont
  {Orignac},\ and\ \citenamefont {Rigol}}]{Cazalilla11}%
  \BibitemOpen
  \bibfield  {author} {\bibinfo {author} {\bibfnamefont {M.~A.}\ \bibnamefont
  {Cazalilla}}, \bibinfo {author} {\bibfnamefont {R.}~\bibnamefont {Citro}},
  \bibinfo {author} {\bibfnamefont {T.}~\bibnamefont {Giamarchi}}, \bibinfo
  {author} {\bibfnamefont {E.}~\bibnamefont {Orignac}}, \ and\ \bibinfo
  {author} {\bibfnamefont {M.}~\bibnamefont {Rigol}},\ }\href {\doibase
  10.1103/RevModPhys.83.1405} {\bibfield  {journal} {\bibinfo  {journal} {Rev.
  Mod. Phys.}\ }\textbf {\bibinfo {volume} {83}},\ \bibinfo {pages} {1405}
  (\bibinfo {year} {2011})}\BibitemShut {NoStop}%
\bibitem [{\citenamefont {Zupancic}\ \emph {et~al.}(2019)\citenamefont
  {Zupancic}, \citenamefont {Dreon}, \citenamefont {Li}, \citenamefont
  {Baumg\"artner}, \citenamefont {Morales}, \citenamefont {Zheng},
  \citenamefont {Cooper}, \citenamefont {Esslinger},\ and\ \citenamefont
  {Donner}}]{Zupancic19}%
  \BibitemOpen
  \bibfield  {author} {\bibinfo {author} {\bibfnamefont {P.}~\bibnamefont
  {Zupancic}}, \bibinfo {author} {\bibfnamefont {D.}~\bibnamefont {Dreon}},
  \bibinfo {author} {\bibfnamefont {X.}~\bibnamefont {Li}}, \bibinfo {author}
  {\bibfnamefont {A.}~\bibnamefont {Baumg\"artner}}, \bibinfo {author}
  {\bibfnamefont {A.}~\bibnamefont {Morales}}, \bibinfo {author} {\bibfnamefont
  {W.}~\bibnamefont {Zheng}}, \bibinfo {author} {\bibfnamefont {N.~R.}\
  \bibnamefont {Cooper}}, \bibinfo {author} {\bibfnamefont {T.}~\bibnamefont
  {Esslinger}}, \ and\ \bibinfo {author} {\bibfnamefont {T.}~\bibnamefont
  {Donner}},\ }\href {\doibase 10.1103/PhysRevLett.123.233601} {\bibfield
  {journal} {\bibinfo  {journal} {Phys. Rev. Lett.}\ }\textbf {\bibinfo
  {volume} {123}},\ \bibinfo {pages} {233601} (\bibinfo {year}
  {2019})}\BibitemShut {NoStop}%
\bibitem [{\citenamefont {Mishra}\ \emph {et~al.}(2016)\citenamefont {Mishra},
  \citenamefont {Greschner},\ and\ \citenamefont {Santos}}]{Mishra16}%
  \BibitemOpen
  \bibfield  {author} {\bibinfo {author} {\bibfnamefont {T.}~\bibnamefont
  {Mishra}}, \bibinfo {author} {\bibfnamefont {S.}~\bibnamefont {Greschner}}, \
  and\ \bibinfo {author} {\bibfnamefont {L.}~\bibnamefont {Santos}},\ }\href
  {\doibase 10.1088/1367-2630/18/4/045016} {\bibfield  {journal} {\bibinfo
  {journal} {New J. Phys.}\ }\textbf {\bibinfo {volume} {18}},\ \bibinfo
  {pages} {045016} (\bibinfo {year} {2016})}\BibitemShut {NoStop}%
\bibitem [{\citenamefont {Di~Dio}\ \emph {et~al.}(2014)\citenamefont {Di~Dio},
  \citenamefont {Barbiero}, \citenamefont {Recati},\ and\ \citenamefont
  {Dalmonte}}]{Dio14}%
  \BibitemOpen
  \bibfield  {author} {\bibinfo {author} {\bibfnamefont {M.}~\bibnamefont
  {Di~Dio}}, \bibinfo {author} {\bibfnamefont {L.}~\bibnamefont {Barbiero}},
  \bibinfo {author} {\bibfnamefont {A.}~\bibnamefont {Recati}}, \ and\ \bibinfo
  {author} {\bibfnamefont {M.}~\bibnamefont {Dalmonte}},\ }\href {\doibase
  10.1103/PhysRevA.90.063608} {\bibfield  {journal} {\bibinfo  {journal} {Phys.
  Rev. A}\ }\textbf {\bibinfo {volume} {90}},\ \bibinfo {pages} {063608}
  (\bibinfo {year} {2014})}\BibitemShut {NoStop}%
\bibitem [{\citenamefont {Zhang}\ \emph {et~al.}(2021)\citenamefont {Zhang},
  \citenamefont {Zhang}, \citenamefont {Yang},\ and\ \citenamefont
  {Capogrosso-Sansone}}]{Zhang21}%
  \BibitemOpen
  \bibfield  {author} {\bibinfo {author} {\bibfnamefont {C.}~\bibnamefont
  {Zhang}}, \bibinfo {author} {\bibfnamefont {J.}~\bibnamefont {Zhang}},
  \bibinfo {author} {\bibfnamefont {J.}~\bibnamefont {Yang}}, \ and\ \bibinfo
  {author} {\bibfnamefont {B.}~\bibnamefont {Capogrosso-Sansone}},\ }\href
  {\doibase 10.1103/PhysRevA.103.043333} {\bibfield  {journal} {\bibinfo
  {journal} {Phys. Rev. A}\ }\textbf {\bibinfo {volume} {103}},\ \bibinfo
  {pages} {043333} (\bibinfo {year} {2021})}\BibitemShut {NoStop}%
\bibitem [{\citenamefont {Fishman}\ \emph {et~al.}()\citenamefont {Fishman},
  \citenamefont {White},\ and\ \citenamefont {Stoudenmire}}]{itensor}%
  \BibitemOpen
  \bibfield  {author} {\bibinfo {author} {\bibfnamefont {M.}~\bibnamefont
  {Fishman}}, \bibinfo {author} {\bibfnamefont {S.~R.}\ \bibnamefont {White}},
  \ and\ \bibinfo {author} {\bibfnamefont {E.~M.}\ \bibnamefont
  {Stoudenmire}},\ }\href@noop {} {\enquote {\bibinfo {title} {The
  \mbox{ITensor} software library for tensor network calculations},}\ }\Eprint
  {http://arxiv.org/abs/2007.14822} {arXiv:2007.14822} \BibitemShut {NoStop}%
\bibitem [{\citenamefont {Sebby-Strabley}\ \emph {et~al.}(2006)\citenamefont
  {Sebby-Strabley}, \citenamefont {Anderlini}, \citenamefont {Jessen},\ and\
  \citenamefont {Porto}}]{Schuetz13}%
  \BibitemOpen
  \bibfield  {author} {\bibinfo {author} {\bibfnamefont {J.}~\bibnamefont
  {Sebby-Strabley}}, \bibinfo {author} {\bibfnamefont {M.}~\bibnamefont
  {Anderlini}}, \bibinfo {author} {\bibfnamefont {P.~S.}\ \bibnamefont
  {Jessen}}, \ and\ \bibinfo {author} {\bibfnamefont {J.~V.}\ \bibnamefont
  {Porto}},\ }\href {\doibase 10.1103/PhysRevA.73.033605} {\bibfield  {journal}
  {\bibinfo  {journal} {Phys. Rev. A}\ }\textbf {\bibinfo {volume} {73}},\
  \bibinfo {pages} {033605} (\bibinfo {year} {2006})}\BibitemShut {NoStop}%
\bibitem [{\citenamefont {Ritsch}\ \emph {et~al.}(2013)\citenamefont {Ritsch},
  \citenamefont {Domokos}, \citenamefont {Brennecke},\ and\ \citenamefont
  {Esslinger}}]{Ritsch13}%
  \BibitemOpen
  \bibfield  {author} {\bibinfo {author} {\bibfnamefont {H.}~\bibnamefont
  {Ritsch}}, \bibinfo {author} {\bibfnamefont {P.}~\bibnamefont {Domokos}},
  \bibinfo {author} {\bibfnamefont {F.}~\bibnamefont {Brennecke}}, \ and\
  \bibinfo {author} {\bibfnamefont {T.}~\bibnamefont {Esslinger}},\ }\href
  {\doibase 10.1103/RevModPhys.85.553} {\bibfield  {journal} {\bibinfo
  {journal} {Rev. Mod. Phys.}\ }\textbf {\bibinfo {volume} {85}},\ \bibinfo
  {pages} {553} (\bibinfo {year} {2013})}\BibitemShut {NoStop}%
\bibitem [{\citenamefont {Singh}\ \emph {et~al.}(2010)\citenamefont {Singh},
  \citenamefont {Pfeifer},\ and\ \citenamefont {Vidal}}]{Singh10}%
  \BibitemOpen
  \bibfield  {author} {\bibinfo {author} {\bibfnamefont {S.}~\bibnamefont
  {Singh}}, \bibinfo {author} {\bibfnamefont {R.~N.~C.}\ \bibnamefont
  {Pfeifer}}, \ and\ \bibinfo {author} {\bibfnamefont {G.}~\bibnamefont
  {Vidal}},\ }\href {\doibase 10.1103/PhysRevA.82.050301} {\bibfield  {journal}
  {\bibinfo  {journal} {Phys. Rev. A}\ }\textbf {\bibinfo {volume} {82}},\
  \bibinfo {pages} {050301} (\bibinfo {year} {2010})}\BibitemShut {NoStop}%
\bibitem [{\citenamefont {Singh}\ \emph {et~al.}(2011)\citenamefont {Singh},
  \citenamefont {Pfeifer},\ and\ \citenamefont {Vidal}}]{Singh11}%
  \BibitemOpen
  \bibfield  {author} {\bibinfo {author} {\bibfnamefont {S.}~\bibnamefont
  {Singh}}, \bibinfo {author} {\bibfnamefont {R.~N.~C.}\ \bibnamefont
  {Pfeifer}}, \ and\ \bibinfo {author} {\bibfnamefont {G.}~\bibnamefont
  {Vidal}},\ }\href {\doibase 10.1103/PhysRevB.83.115125} {\bibfield  {journal}
  {\bibinfo  {journal} {Phys. Rev. B}\ }\textbf {\bibinfo {volume} {83}},\
  \bibinfo {pages} {115125} (\bibinfo {year} {2011})}\BibitemShut {NoStop}%
\bibitem [{\citenamefont {Crosswhite}\ \emph {et~al.}(2008)\citenamefont
  {Crosswhite}, \citenamefont {Doherty},\ and\ \citenamefont
  {Vidal}}]{Crosswhite08}%
  \BibitemOpen
  \bibfield  {author} {\bibinfo {author} {\bibfnamefont {G.~M.}\ \bibnamefont
  {Crosswhite}}, \bibinfo {author} {\bibfnamefont {A.~C.}\ \bibnamefont
  {Doherty}}, \ and\ \bibinfo {author} {\bibfnamefont {G.}~\bibnamefont
  {Vidal}},\ }\href {\doibase 10.1103/PhysRevB.78.035116} {\bibfield  {journal}
  {\bibinfo  {journal} {Phys. Rev. B}\ }\textbf {\bibinfo {volume} {78}},\
  \bibinfo {pages} {035116} (\bibinfo {year} {2008})}\BibitemShut {NoStop}%
\bibitem [{\citenamefont {Pirvu}\ \emph {et~al.}(2010)\citenamefont {Pirvu},
  \citenamefont {Murg}, \citenamefont {Cirac},\ and\ \citenamefont
  {Verstraete}}]{Pirvu10}%
  \BibitemOpen
  \bibfield  {author} {\bibinfo {author} {\bibfnamefont {B.}~\bibnamefont
  {Pirvu}}, \bibinfo {author} {\bibfnamefont {V.}~\bibnamefont {Murg}},
  \bibinfo {author} {\bibfnamefont {J.~I.}\ \bibnamefont {Cirac}}, \ and\
  \bibinfo {author} {\bibfnamefont {F.}~\bibnamefont {Verstraete}},\ }\href
  {\doibase 10.1088/1367-2630/12/2/025012} {\bibfield  {journal} {\bibinfo
  {journal} {New Journal of Physics}\ }\textbf {\bibinfo {volume} {12}},\
  \bibinfo {pages} {025012} (\bibinfo {year} {2010})}\BibitemShut {NoStop}%
\end{thebibliography}%

\end{document}